# New Results on the Thermodynamical Properties of the Climate System

*Valerio Lucarini* [v.lucarini@reading.ac.uk]

Department of Meteorology & Department of Mathematics, University of Reading, Reading, UK

*Klaus Fraedrich & Francesco Ragone*

Meteorologisches Institut, Klimacampus, University of Hamburg, Grindelberg 5, Hamburg, Germany

## Abstract

In this paper we exploit two equivalent formulations of the average rate of material entropy production in the climate system to propose an approximate splitting between contributions due to vertical and eminently horizontal processes. Our approach is based only upon 2D radiative fields at the surface and at the top of atmosphere. Using 2D fields at the top of atmosphere alone, we derive lower bounds to the rate of material entropy production and to the intensity of the Lorenz energy cycle. By introducing a measure of the efficiency of the planetary system with respect to horizontal thermodynamical processes, we provide insight on a previous intuition on the possibility of defining a baroclinic heat engine extracting work from the meridional heat flux. The approximate formula of the material entropy production is verified and used for studying the global thermodynamic properties of climate models (CMs) included in the PCMDI/CMIP3 dataset in pre-industrial climate conditions. It is found that about 90% of the material entropy production is due to vertical processes such as convection, whereas the large scale meridional heat transport contributes only about 10%. This suggest that the traditional 2-box models used for providing a minimal representation of entropy production in planetary systems are not appropriate, while a basic – but conceptually correct – description can be framed in terms of a 4-box model. The total material entropy production is typically 55 mWK$^{-1}$m$^{-2}$, with discrepancies of the order of 5% and CMs' baroclinic efficiencies are clustered around 0.055. The lower bounds on the intensity of the Lorenz energy cycle featured by CMs are found to be around 1.0-1.5Wm$^{-2}$, which implies that the derived inequality is rather stringent. When looking at the variability and co-variability of the considered thermodynamical quantities, the agreement among CMs is worse, suggesting that the description of feedbacks is more uncertain. The contributions to material entropy production from vertical and horizontal processes are positively correlated, so that no compensation mechanism seems in place. Quite consistently among CMs, the variability of the efficiency of the system is a better proxy for variability of the entropy production due to horizontal processes than that of the large scale heat flux. The possibility of providing constraints to the 3D dynamics of the fluid envelope based only upon 2D observations of radiative fluxes seems promising for the observational study of planets and for testing numerical models.

## 1.   Introduction

It has long been recognized that adopting a thermodynamical perspective, as pioneered by Lorenz (1955, 1967), may prove of great utility for providing a satisfactory theory of climate dynamics able to tackle simultaneously balances of physical quantities and dynamical instabilities, and aimed at explaining the global structural properties of the climate system, as envisaged by Saltzman (2002). This is of great relevance in terms of the pursuit for explaining climate variability and change on a large variety of scales, covering major paleoclimatic shifts, almost regularly repeated events such as ice ages, as well as the ongoing and future anthropogenic climate change. Additionally, this strategy may prove of great relevance for the provision of reliable metrics for the validation of climate models, as asked for by the Intergovernmental Panel on Climate Change (IPCC 2007) and discussed, e.g., in Held (2005), Lucarini (2008a). See Lucarini and Ragone (2010) for a recent example.

Along the lines of non-equilibrium macroscopic thermodynamics (Prigogine 1962, De Groot and Mazur 1984), the climate can be seen as a non-equilibrium system, which transforms potential into mechanical energy like a thermal engine and generates entropy by irreversible processes. When the external and internal parameters have fixed values, the climate system achieves a steady state by balancing the thermodynamical fluxes with the surrounding environment (Peixoto and Oort 1992).

The concept of the energy cycle of the atmosphere introduced by Lorenz (1955, 1967) allowed for defining an effective climate machine such that the atmospheric and oceanic motions simultaneously result from the mechanical work (then dissipated in a turbulent cascade) produced by the engine, and re-equilibrate the energy balance of the climate system (Stone 1978a,b, Barry et al. 2002). Johnson (2000) introduced a Carnot engine–equivalent picture of the climate system by defining effective warm and the cold reservoirs and their temperatures. Recently, Tailleux (2009) proposed a fresh outlook on the energetics of the ocean circulation along similar lines.

The interest towards studying the climate irreversibility largely stems from the proposal of the maximum entropy production principle (MEPP) by Paltridge (1975), which suggests that the climate is a non-equilibrium nonlinear systems adjusting in such a way to maximize the entropy production (Grassl 1981, Mobbs 1982, Kleidon and Lorenz 2005). Even if recent claims of *ab-initio* derivation of MEPP (Dewar 2005) have been dismissed (Grinstein and Linsker 2007), strong criticisms have arisen within the geophysical community (Goody 2007), this principle has stimulated the re-examination of entropy production in the climate system (Peixoto et al. 1991, Goody 2000, Fraedrich and Lunkeit 2008, Pascale et al. 2009) and the development of new strategies for improving the parameterisations of climate models (Kleidon et al. 2006). Moreover, a more detailed analysis of the various processes responsible for the entropy production has lead to clarifying the relative role of contributions due to radiative processes, mainly related to the degradation of the photons exergy by the thermalisation of the solar radiation at terrestrial temperatures (Wu and Liu 2009), and those due to the turbulent processes related to the motions of the fluid envelope of the planet (Goody 2000). The latter contributions, which add up to the so-called material entropy production, albeit relatively small, are expected to be of greater relevance as diagnostics of the large scale properties of the system (Ozawa et al. 2003, Lucarini 2009).

In recent years, a great effort has been devoted to improving the thermodynamical description of the climate system by taking into account in a more profound way the irreversible processes associated with the mixing and phase changes of water vapour (Goody 2000, Pauluis 2000, Pauluis and Held 2002a,b, Romps 2008). Whilst these theoretical contributions are crucial for assessing with a higher degree of precision the components of the entropy budget of the climate system, we shall see that a simpler and more manageable description of the climate thermodynamics has excellent performances for estimating the material entropy production of the system.

We should note that when the material entropy production of the climate system is analysed, most of the attention is devoted to atmospheric processes, with the ocean being of relevance only as boundary conditions with a role not qualitatively different from that of land surface. While being of great relevance when the energetics of the climate system is considered, the contribution to the material entropy production resulting from ocean processes is indeed negligible (Shinokawa and Ozawa 2001), basically because, apart from the mixed layer, the world ocean is up to a good approximation an isothermal fluid. The oceanic entropy production is found to be less than 2% of the atmospheric one (Pascale et al. 2009).

Recently, a link has been found between the Carnot efficiency, the intensity of the Lorenz energy cycle, the material entropy production and the degree of irreversibility of the climate system (Lucarini 2009). Namely, the efficiency of the equivalent thermal machine sets also the proportionality between the internal entropy fluctuation of the system and the lower bound to entropy production by the fluid compatible with the $2^{nd}$ law of thermodynamics. Such a bound is basically given by the entropy produced by the dissipation of the mechanical energy, whereas the excess of entropy production is due to the turbulent transport of heat down the gradient the temperature field.

These results have paved the way for a new, extensive exploration aimed at understanding the climate response under various scenarios of forcings, of atmospheric composition, and of boundary conditions. Recent preliminary efforts using PLASIM (Fraedrich et al. 2005), a simplified yet Earth-like climate model, have focused on the impacts on the thermodynamics of the climate system of changes in the solar constant, with the analysis of the onset and decay of snowball Earth conditions (Lucarini et al. 2010a), and on those due to changing $CO_2$ concentration (Lucarini et al. 2010b).

When performing a complete thermodynamic analysis of the planetary system a potentially serious problem is that three dimensional, time dependent information on the intensive thermodynamical quantities, of their tendencies, and of the forcing terms are required (Lucarini 2009). It has been proved that for a given numerical model constructing the suitable diagnostic tools is feasible and definitely not computationally burdensome, as only integral operators (which boil down to weighted sums) are involved (Fraedrich and Lunkeit 2008). On the other side, deducing the thermodynamics of the system from observations is a rather complex matter, since it is hard to obtain accurate reconstructions of all the involved 3D thermodynamic quantities with sufficient spatial and temporal resolution. Such a problem arises also when an intercomparison analysis covering a large number of CMs is pursued, because the needed climate fields are not necessarily among those routinely provided for public download, as discussed by Lucarini and Ragone (2011).

In this paper we introduce an approximate formula which allows for splitting the material entropy production into two contributions, one related to horizontal, planetary scale transport processes, the other one related to vertical processes. The two contributions can be computed separately using 2D radiative fields at the top of the atmosphere (TOA) and at surface. The contribution to the material entropy production due to horizontal processes can be computed using 2D radiative fields at the top of the atmosphere only. Such estimate allows for computing a lower bound to the total values of material entropy production and intensity of the Lorenz energy cycle, or, equivalently, of the average rate of total dissipation of the kinetic energy. Moreover, by combining energy and entropy constraints, it is possible to provide insights on a previous intuition on the possibility of defining a baroclinic heat engine extracting work from the meridional heat flux (Barry et al. 2002). We take advantage of our approach to analyse the thermodynamic properties of the pre-industrial (PI) control runs of the CMs included in the PCMDI/CMIP3 dataset, thus extending the results of Lucarini and Ragone (2011) to the "$2^{nd}$ law of thermodynamics" diagnostics. We take into account first and second moments of thermodynamical quantities describing out-of-equilibrium properties, in order to assess their climatology and their (co-)variability, thus testing equilibration processes.

The paper is divided as follows. In Section 2 we review two distinct formulas allowing for evaluating the material entropy production of a generic planetary system and perform a suitable scale analysis, obtaining simplified expressions making use of 2D fields at the surface and TOA only. In section 3 we show how to derive from the above described approximate formulas general bounds on the material entropy production and on the degree of irreversibility of the system. We also clarify the basic features required in simplified box-models (Kleidon and Lorenz 2005) for describing the fundamental properties of entropy production in the Earth system. In Section 4 we verify the validity of the approximate formula and exploit it to estimate the thermodynamic properties of the CMs included in the PCMDI/CMIP3 dataset in preindustrial climate conditions. In Section 5 we present our conclusions and perspectives for future work.

## 2. Rate of Material Entropy Production

*2.1 Theoretical Outlook*

The traditional approach for the investigation of the entropy production of the climate system relies on separating the contributions due to irreversible processes involving matter and those due to irreversible changes in the spectral properties of the radiation.

The process of thermalisation of the solar radiation gives by far the most important contribution to the global planetary entropy production, basically as it involves the transformation of electromagnetic energy travelling through space obeying a Planckian spectrum with the temperature signature of the Sun's corona (about 5800 K) into (quantitatively identical) electromagnetic energy whose spectral properties are approximately described by a Planckian spectrum at the Earth's emission temperature (about 250 K). A very detailed and extensive account of these processes and their contribution in terms of entropy production has recently been given by Wu and Liu (2009).

The rest of the irreversible processes taking place in the climate system provide a much smaller contribution to the entropy production, basically because much less relevant temperature (or chemical potential) differences are involved. The irreversible transformations occurring in the climate systems involve in principle a great variety of phenomena, including dissipation of mechanical energy, heat transport down the temperature gradient, irreversible mixing and phase transitions.

In a system at steady state, the expectation value of the extensive, integrated material entropy does not depend on time. Following (Johnson 2000; Goody 2000), it is possible to derive the following equation for the total entropy $S$ of the climate system:

$$\overline{\dot{S}} = \int_\Omega \left( \overline{\frac{\dot{q}_{rad}}{T}} + \overline{\dot{s}_{turb}} \right) dV = 0. \tag{1}$$

where $\Omega$ is the spatial domain of integration, the dot indicates the operation of time derivative, the overbar indicates the operation of long term average, $T$ is the local temperature of the medium, $\dot{q}_{rad}$ is the heating due to the convergence of the radiation fluxes, while $\dot{s}_{turb}$ is the density of entropy production due to irreversible processes involving the fluid medium, and is usually referred to as local material entropy production (Ozawa et al. 2003). Therefore, we obtain:

$$\overline{\dot{S}_{mat}} = \int_\Omega \overline{\dot{s}_{turb}} dV = -\int_\Omega \overline{\frac{\dot{q}_{rad}}{T}} dV \qquad (2)$$

where $\overline{\dot{S}_{mat}}$ is the average rate of material entropy production of the climate system. This equation provides us with two recipes for computing the material entropy production, one based on the direct computation of the spatial integral of $\dot{s}_{turb}$, and the other (the "inverse formula") based on the evaluation of the interaction between radiation and matter. Depending on the degree of detail and precision we adopt in the representation of the physicochemical properties of the climate system, we may derive different expressions for $\dot{s}_{turb}$, which are suited for characterizing a wider or smaller number of irreversible processes.

Commonly, the following expression is adopted for the computation of the material entropy production budget:

$$\dot{s}_{turb} = \frac{\varepsilon^2}{T} + \vec{F}_{SH} \cdot \vec{\nabla} \frac{1}{T} + \vec{F}_{LH} \cdot \vec{\nabla} \frac{1}{T}. \qquad (3)$$

In this expression, $\varepsilon^2$ is the dissipation of kinetic energy, which also takes into account the friction of the falling hydrometeors - see discussion on this rather controversial issue in (Goody 2000, Pauluis et al. 2000, Rennó 2001, Lorenz and Rennó 2002, Kleidon and Lorenz 2005), while $\vec{F}_{SH}$ and $\vec{F}_{LH}$ are the sensible and latent heat fluxes, respectively. Note that $\vec{F}_{SH}$ refers to a small-scale turbulent flux, while $\vec{F}_{LH}$ refers to both small-scale turbulent and large scale fluxes, the large-scale transport of water vapor being an important component of the global entropy production budget. The hydrological cycle is indeed a key element of the thermodynamics of the climate system. Such a formula has been widely used for estimating the entropy production of the Earth system (Peixoto et al 1991, Peixoto and Oort 1992, Fraedrich and Lunkeit 2008, Pascale et al. 2009, Lucarini et al. 2010a, 2010b)

Romps (2008) refers to the representation of the entropy production given by Eq. (3) as resulting from the bulk heating budget, because water is treated mainly as a passive substance, while processes such as irreversible mixing of the water vapour is altogether ignored. More detailed descriptions of the "moist" atmosphere have led to formulations of atmospheric thermodynamics able to account for these processes (Goody 2000, Pauluis 2000, Pauluis and Held 2002a,b, Romps 2008). Moreover, additional contributions to the entropy production are given by the irreversible mixing of salt in the ocean (Shinokawa and Ozawa 2001). Such more refined formulations of the entropy processes inside the climate system account for a consistent treatment of the entropy generated by the hydrological cycle. Given the complex nature of the climate system, additional processes contributing to entropy production can be highlighted. Kleidon (2009) presents a complete and holistic account of this issue, suggesting that biological and geochemical processes, as well as the great variety of chemical processes taking place in the atmosphere should in principle be included to get a truly complete treatment of the entropy budget of the Earth system. None of these processes are usually explicitly represented in current climate models.

Since we aim at a parsimonious but efficient representation of the entropy production of the climate system, we would like to be able to choose an approach, which translates into an explicit expression for $\dot{s}_{turb}$, which is as simple as possible but, at the same time, provides a good approximation to the entropy production. Moreover, when using our formula to compute the entropy production using data from a numerical model, it makes no sense to take into account processes which are not represented in the model. One should also note that the entropy production due to unphysical processes present in the model (numerical diffusion, dispersion, etc..) should be in principle taken into account if accurate calculations are sought after. Previous analyses of the detailed budget of entropy production in climate models have shown that these contributions are, fortunately, almost negligible when the total entropy production is considered, and not the various contributions coming from the various involved irreversible physical processes (Pascale et. al 2009, see below).

On the other hand, the time average of the volume integration of $-\dot{q}_{rad}/T$ gives the exact value of the entropy production, so that the indirect approach allows on one side to obtain precise estimates, and on the other side to test the validity of different explicit formulations for $\dot{s}_{turb}$ and, consequently, of the level of detail we adopt in describing the climatic thermodynamic processes.

Interpreting previous climate models' results is definitely useful in this direction. Pascale et al. (2009), who considered two fully coupled atmosphere-ocean climate models, clearly showed that in present day conditions the identity given by Eq. (2) is obeyed up to an excellent degree of precision when we use for $\dot{s}_{turb}$ the expression given in Eq. (3). The observed agreement is within 1% and of the same order of magnitude of the uncertainty in the material entropy production due to the bias in the imperfect closure of the energy cycle due to spurious energy sinks/sources inside the system (Lucarini et al. 2010a,b; Lucarini and Ragone 2011). We may then conclude that whereas the sophisticated thermodynamical framework developed in (Pauluis 2000, Pauluis and Held 2002a,b, Romps 2008) is crucial for understanding in detail the various terms contributing to the entropy budget of the climate system, the simpler formulation (Peixoto et al 1991, Peixoto and Oort 1992, Fraedrich and Lunkeit 2008, Pascale et al. 2009, Lucarini et al. 2010a, 2010b) provides rather accurate results when only the material entropy production is taken into account. Note that Goody (2000) observed that adopting a more detailed physical description of the atmosphere where irreversible water vapour mixing processes are considered changes only slightly the estimate of the material entropy production. He attributed this to a spurious coincidence, but given the Pascale et al. (2009)'s results, it is way more likely that treating the latent heat release and absorption in a simplified way boils down to adding up the most important contributions composing the detailed entropy budget advocated in (Pauluis 2000, Pauluis and Held 2002a,b, Romps 2008).

It has also to be noted that the overall contribution of the subsurface oceanic processes to the entropy production of the climate system is negligible, the basic reason being that to a good approximation the deep ocean is an isothermal fluid, so that heat transport generates little entropy (Pascale et al. 2009). From now on, we consider entropy generation as a matter involving only processes in the atmosphere and processes entailing energy exchanges between the atmosphere and the underlying surface, which can correspond to either land or ocean.

*2.2 Scale Analysis: Direct Expression of Entropy Production*

The "direct" expression for the material entropy production is considered first. Using Gauss' theorem, we can write:

$$\overline{\dot{S}_{mat}} = \int_\Omega \overline{\frac{\varepsilon^2}{T}} dV + \int_\Omega \overline{\vec{F} \cdot \vec{\nabla} \frac{1}{T}} dV = \int_\Omega \overline{\frac{\varepsilon^2}{T}} dV + \int_\Omega \overline{\vec{\nabla} \cdot \left(\frac{\vec{F}}{T}\right)} dV - \int_\Omega \overline{\left(\frac{\vec{\nabla} \cdot \vec{F}}{T}\right)} dV =$$

$$= \int_\Omega \overline{\frac{\varepsilon^2}{T}} dV + \int_{\partial\Omega} \overline{\hat{n} \cdot \left(\frac{\vec{F}}{T}\right)} d\sigma - \int_\Omega \overline{\left(\frac{\vec{\nabla} \cdot \vec{F}}{T}\right)} dV = \int_\Omega \overline{\frac{\varepsilon^2}{T}} dV - \int_\Omega \overline{\left(\frac{\vec{\nabla} \cdot \vec{F}}{T}\right)} dV . \qquad (4)$$

As discussed in Lucarini (2009a), the volume integral of the first term on the right hand side of Eq. (4) basically gives us the lower bound to the entropy production compatible with the system having a global mean dissipation of kinetic energy (and production of mechanical work) equal to $\overline{W} = \overline{D} = \int \overline{\varepsilon^2} dV$. The second term in Eq. (4) vanishes as no material fluxes are present at the top of the atmosphere. The material flux $\vec{F}$ accounts for both the sensible and latent heat fluxes, so that $\vec{F} = \vec{F}_{SH} + \vec{F}_{LH}$. The two fluxes have rather different properties since sensible heat turbulent transport is mostly relevant for the interaction between the surface and the boundary layer of the atmosphere, whereas the latent heat flux is such that it picks up water vapour at unsaturated, relatively warm surface conditions and transports it to regions with lower temperatures where condensation and precipitation occur. Turbulent transport occurs mainly in the vertical direction (Peixoto and Oort 1992), so that $\vec{\nabla} \cdot \vec{F}_{SH} \approx \partial_z F_z^{SH}$. In the case of latent heat, this approximation does not hold, since the large-scale horizontal transport is an important component of $\vec{F}_{LH}$. Nevertheless, extending the approach by Fraedrich and Lunkeit (2008), we can rewrite Eq. (4) in a simpler 2D form as

$$\overline{\dot{S}_{mat}} = \overline{\dot{S}_{mat}^{diss}} + \overline{\dot{S}_{mat}^{SH}} + \overline{\dot{S}_{mat}^{LH}} = \int_A \frac{\langle\overline{\varepsilon^2}\rangle}{T_{diss}} d\sigma - \int_A \overline{F_{z,surf}^{SH}} \left(\frac{1}{T_{SH}^+} - \frac{1}{T_{SH}^-}\right) d\sigma - \int_A \overline{F_{z,surf}^{LH}} \left(\frac{1}{T_{LH}^+} - \frac{1}{T_{LH}^-}\right) d\sigma + \Delta \overline{\dot{S}_{mat}^{LH}} \quad (5)$$

where $\overline{\dot{S}_{mat}^{diss}}$, $\overline{\dot{S}_{mat}^{SH}}$ and $\overline{\dot{S}_{mat}^{LH}}$ are respectively the contributions to the material entropy production due to energy dissipation, sensible and latent heat fluxes, $\langle\overline{\varepsilon^2}\rangle = \int \overline{\varepsilon^2} dz$ is the vertically integrated dissipational heating, $T_{diss} = \int \overline{\varepsilon^2} dz / \int \overline{\varepsilon^2/T} dz$ is a suitably time averaged 2D field of characteristic temperature for energy dissipation, $\overline{F_{z,surf}^{SH}}$ and $\overline{F_{z,surf}^{LH}}$ are the long term averages of the vertical components of sensible and latent heat fluxes at the surface (positive upwards), $\overline{T_{SH}^+}$, $\overline{T_{LH}^+}$, $\overline{T_{SH}^-}$ and $\overline{T_{LH}^-}$ are 2D fields of locally (in each column) defined characteristic temperatures respectively of removal and absorption of sensible and latent heat, and $\Delta \overline{\dot{S}_{mat}^{LH}}$ is a correction term whose meaning is explained in the following. Formally, this is achieved by dividing each column into two domains in the z-direction, one where $\partial_z F_z$ is positive, and one where $\partial_z F_z$ is negative. The two temperatures are then defined similarly to $T_{diss}$. Note that this operation is performed separately for the latent and sensible heat flux. As in each column most of the sensible and latent heat is removed at or very close to the surface, we assume $T_{SH}^+ \approx T_{LH}^+ \approx T_S$. Instead, we expect that $T_{SH}^-$ is closely approximated by the temperature of the boundary layer $T_{BL}$, while we indicate $T_{LH}^-$ with $T_C$, since it refers to the average condensation temperature. Therefore we can write

$$\overline{\dot{S}_{mat}} = \overline{\dot{S}_{mat}^{diss}} + \overline{\dot{S}_{mat}^{SH}} + \overline{\dot{S}_{mat}^{LH}} = \int_A \frac{\langle \varepsilon^2 \rangle}{T_{diss}} d\sigma - \int_A \overline{F_{z,surf}^{SH}} \left( \frac{1}{T_S} - \frac{1}{T_{BL}} \right) d\sigma - \int_A \overline{F_{z,surf}^{LH}} \left( \frac{1}{T_S} - \frac{1}{T_C} \right) d\sigma + \Delta \overline{\dot{S}_{mat}^{LH}} \qquad (6)$$

Let's now see how the correction term $\Delta \overline{\dot{S}_{mat}^{LH}}$ arises. In the case of the latent heat, because of the presence of horizontal large-scale transports of moisture, it is not possible to define rigorously the condensation temperatures locally in each column, since not all the moisture which evaporates at the bottom of a certain atmospheric column will condensate right above, partly being transported horizontally outside of the column. In present day climate moisture is transported from the tropical regions between 10° and 40° N (S) to the equatorial region and to the mid-high latitudes, fuelling respectively the deep convection in the uplifting branch of the Hadley cell and the baroclinic processes of the midlatitudes (Peixoto and Oort 1992, table 7.1). We can write the latent heat due to surface evaporation as $\overline{F_{z,surf}^{LH}} = \overline{F_{loc}^{LH}} + \overline{F_{out}^{LH}}$, where $\overline{F_{loc}^{LH}}$ and $\overline{F_{out}^{LH}}$ are the latent heat fluxes related respectively to the the part of water vapor which condenses at the condensation temperature $T_C$ in the same column where it is evaporated, and to the part which is transported and condensed somewhere else at the effective condensation temperature $T_C'$. The correct formulation for the material entropy production related to the phase changes of the water evaporated in a certain atmospheric column is therefore (neglecting contributions from re-evaporation of falling rain droplets):

$$\overline{\dot{s}_{mat}^{LH}} = -\overline{F_{loc}^{LH}}\left(\frac{1}{T_S}-\frac{1}{T_C}\right) - \overline{F_{out}^{LH}}\left(\frac{1}{T_S}-\frac{1}{T_C'}\right) \qquad (7)$$

We can rewrite Eq. (7) as:

$$\overline{\dot{s}_{mat}^{LH}} = -\left(\overline{F_{loc}^{LH}}+\overline{F_{out}^{LH}}\right)\left(\frac{1}{T_S}-\frac{1}{T_C}\right) - \overline{F_{out}^{LH}}\left(\frac{1}{T_C}-\frac{1}{T_C'}\right) = -\overline{F_{z,surf}^{LH}}\left(\frac{1}{T_S}-\frac{1}{T_C}\right) + \Delta\overline{\dot{s}_{mat}^{LH}} \qquad (8)$$

where:

$$\Delta\overline{\dot{s}_{mat}^{LH}} = -\overline{F_{out}^{LH}}\left(\frac{1}{T_C}-\frac{1}{T_C'}\right) \qquad (9)$$

is the correction term due to the fact that part of the moisture which evaporates at a certain location is transported and condensed elsewhere. If the effective condensation temperature $T_C'$ is higher than the condensation temperature $T_C$ we have a negative correction term (e.g. moisture transport from the tropics to the equator), positive otherwise (e.g. moisture transport from the tropics to the midlatitudes). It is possible to show that the global integral $\Delta\overline{\dot{S}_{mat}^{LH}}$ of the correction given in Eq (9) is rather small, since its absolute value can be given the upper bound $\Delta\overline{\dot{S}_{mat}^{LH}} < 5\cdot 10^{-3} Wm^{-2}K^{-1}$, one order of magnitude less than the value for $\overline{\dot{S}_{mat}^{LH}}$ computed by Pascale et. al. (2009). We present a detailed derivation of this result in Appendix A.

Since the turbulent transport of sensible heat is an eminently local process, its contribution $\overline{\dot{S}_{mat}^{SH}}$

to the material entropy production is not very large. Instead, the contribution to the material entropy production coming from latent heat ($\overline{\dot{S}_{mat}^{LH}}$) is largely dominant, and involves evaporation, condensation, and transport processes occurring in the context of the hydrological cycle (see also Pascale et al. 2009).

Note that it is possible to define the degree of irreversibility of the system (Lucarini 2009) by introducing the parameter $\alpha = \left(\overline{\dot{S}_{mat}^{SH}} + \overline{\dot{S}_{mat}^{LH}}\right)/\overline{\dot{S}_{mat}^{diss}}$ where symbols refer to Eq. (5). When $\alpha = 0$ the system features the smallest rate of material entropy production compatible with the presence of a Lorenz energy cycle of intensity $\overline{W} = \overline{D} = \int \overline{\varepsilon^2} dV$. If $\alpha = 0$, all the entropy is generated via dissipation of kinetic energy, with no contributions coming from fluxes transporting heat down the gradient of the temperature. Recent model simulations have shown that warmer climate conditions trigger a fast increase in the degree of irreversibility of the Earth system, the main reason being the large sensitivity of latent heat fluxes to increases in the atmospheric and surface temperature (Lucarini et al. 2010a, 2010b).

*2.3 Scale Analysis: Indirect Expression of Entropy Production*

The "indirect" expression for the material entropy production is considered now. We emphasize that such an approach bypasses the problems related to the details in the representation of the atmospheric processes discussed at the beginning of this section. The solar shortwave (SW) radiation heats the system, whereas the infrared longwave (LW) radiation acts as a net cooler. Moreover, we split processes occurring within the atmosphere from those occurring at surface, which we interpret as boundary between the gaseous medium and the solid and liquid medium, thus moving along the lines of the *net exchange formulation* (Green 1967). Along the same lines followed to derive Eq. (5), we write the average rate of material entropy production as follows:

$$\overline{\dot{S}_{mat}} = \int_A \frac{\overline{F_{z,surf}^{SW}} + \overline{F_{z,surf}^{LW}}}{T_S} d\sigma + \int_A \frac{\overline{F_{z,TOA}^{SW}} - \overline{F_{z,surf}^{SW}}}{T_{A,SW}} d\sigma + \int_A \frac{\overline{F_{z,TOA}^{LW}} - \overline{F_{z,surf}^{LW}}}{T_{A,LW}} d\sigma, \qquad (10)$$

where $\overline{F_{z,surf}^{SW}}$ and $\overline{F_{z,surf}^{LW}}$ are the average vertical fluxes of SW and LW radiation at surface, while $\overline{F_{z,TOA}^{SW}}$ and $\overline{F_{z,TOA}^{LW}}$ are the average vertical fluxes of SW and LW radiation at the top of the atmosphere. All the vertical fluxes are considered to be positive upwards, as for the sensible and latent heat fluxes in the previous subsection. Similarly to what discussed in the previous subsection, $T_{A,SW}$ and $T_{A,LW}$ represent the 2D fields of characteristic atmospheric temperatures at which the interactions between matter and SW and LW, respectively, occur. On our planet, most of the SW is absorbed at surface or in the first few meters of ocean, so that the atmosphere is heated from below. We may rewrite Eq. (10) as follows:

$$\overline{\dot{S}_{mat}} = \int_A \overline{F_{z,surf}^{SW}} \left( \frac{1}{T_S} - \frac{1}{T_{A,SW}} \right) d\sigma + \int_A \overline{F_{z,surf}^{LW}} \left( \frac{1}{T_S} - \frac{1}{T_{A,LW}} \right) d\sigma + \int_A \frac{\overline{F_{z,TOA}^{SW}}}{T_{A,SW}} d\sigma + \int_A \frac{\overline{F_{z,TOA}^{LW}}}{T_{A,LW}} d\sigma. \qquad (11)$$

We now assume as an *ansatz* that $T_{A,SW} \approx T_{A,LW} \approx T_E = \sqrt[4]{\overline{F_{z,TOA}^{LW}}/\sigma}$, where $T_E$ is the 2D field of the emission temperature of the planet ($\sigma$ being the Stefan-Boltzmann constant). This means that the vertically averaged characteristic temperature of *atmospheric* absorption and emission are similar. This is the crucial approximation of our approach, motivated by the fact that most of the shortwave absorption in the atmosphere is due to water vapour (Kiehl and Trenberth 1997), which plays a major role also in determining the absorption and emission of longwave radiation in the troposphere. The contribution to shortwave radiation absorption of water vapour is about three times as much as that of $O_3$. This approximation is supported by the estimates presented by Fraedrich and Lunkeit (2008); its validity will be investigated in more quantitative terms in a future publication. For the purposes of the present paper, the approximation will be tested by comparing the values obtained with formula (12) with "exact" values obtained from the direct expression, in the case of the HadCM3 general circulation model analysed in Pascale et al. (2009) (see section 4). We then obtain:

$$\overline{\dot{S}_{mat}} \approx \int_A \left(\overline{F^{SW}_{z,surf}} + \overline{F^{LW}_{z,surf}}\right)\left(\frac{1}{T_S} - \frac{1}{T_E}\right)d\sigma + \int_A \frac{\overline{F^{SW}_{z,TOA}} + \overline{F^{LW}_{z,TOA}}}{T_E}d\sigma$$

$$\approx \int_A \left(\overline{F^{SW}_{z,surf}} + \overline{F^{LW}_{z,surf}}\right)\left(\frac{1}{T_S} - \frac{1}{T_E}\right)d\sigma - \int_A \frac{\overline{\vec{\nabla}_H \cdot \vec{H}}}{T_E}d\sigma$$

$$\approx \overline{\dot{S}^{vert}_{mat}} + \overline{\dot{S}^{hor}_{mat}}, \tag{12}$$

where in the last passage we have used that the convergence of large scale, non-turbulent reversible enthalpy horizontal transport $\vec{H}$ balances the net radiative balance at the top of the atmosphere when long term averages are considered (Peixoto and Oort 1992, Lucarini and Ragone 2011). Equation (12) tells us that the material entropy production can be, alternatively to the splitting proposed in Eq. (5), conceptually decomposed into two terms. The term $\overline{\dot{S}^{vert}_{mat}}$ describes the vertical transport of radiation between two reservoirs, one at the surface temperature, the other one at temperature of the bulk of the atmosphere, and is closely related to dry and moist convective processes (Emanuel 2000). This term treats the fluid envelope as a collection of independent vertical columns dominated by fast exchanges and interactions. The term $\overline{\dot{S}^{hor}_{mat}}$ describes the effect of horizontally transporting energy in a 2D fluid system with spatially varying temperature structure, and is associated to longer time scales. Both terms are positive, the first because the atmosphere is on the average colder than the underlying surface, as the system is heated from below, the second because temperature is expected to be on the average lower where there is convergence of enthalpy fluxes, and larger where divergence is observed, in agreement with the second law of thermodynamics (Peixoto and Oort 1992, Ambaum 2010). The two spatial fields whose integrals give $\overline{\dot{S}_{mat}^{vert}}$ and $\overline{\dot{S}_{mat}^{hor}}$ are expected to have different properties, as the former should be positive everywhere, because each column is described in the first approximation as a quasi-isolated system producing entropy, whereas the sign of the latter will depend on the sign of the radiative budget at TOA.

Note that a discretised version of the second term in Eq. (12), based upon a 2-box approximation of the fluid envelope of the climate system, has been considered as proxy for the total material entropy production (Lorenz et al. 2001, Kleidon 2009). Instead, from Eq. (12) it is apparent that a minimal model of the whole material entropy production of a planetary system must include, in addition to the box of the warm (Box 1) and cold (Box 2) portions of the fluid envelope of the planet, two additional boxes, each representing the planetary surface in the warm (Box 3) and cold (Box 4) regions of the planet. See Fig. 1 for a conceptual scheme, where the arrows indicate the couplings defining the time evolution of the system. Note that the ocean has a dual role: on one side, it acts a lower surface exchanging with the atmosphere sensible and latent heat fluxes (as done by the land surface), on the other side, it has the dynamic role of contributing, as part of the fluid envelope of the planet, to the transport of heat from the warm to the cold box. We remark again that, the internal oceanic processes give only a minor contribution to the entropy production (Pascale et al. 2009).

## 3. Bounds to the Thermodynamical Properties of the Climate System

We now wish to derive some bounds to the thermodynamical properties of the climate system based upon only a very limited set of coarse resolution data. Since the water vapour saturation mixing ratio strongly increases with temperature, and since on Earth the atmospheric temperature decreases with height, the vertical scale of water vapour in globally saturated conditions is smaller than that of the atmosphere. Therefore, we can assume $T_C > T_E$, as confirmed by actual estimates (Peixoto and Oort 1992, Fraedrich and Lunkeit 2008, Pascale et al. 2009). Since, as discussed above, $T_{BL} \approx T_S$, we definitely have that $T_{BL} > T_E$. Therefore, by substituting $T_E$ to the values of the temperatures at which the latent and sensible heat are absorbed, respectively $T_C$ and $T_{BL}$, to the various terms given in Eq. (5), we derive the following inequality:

$$\overline{\dot{S}_{mat}} < \int_A \frac{\overline{\varepsilon^2}}{T_{diss}} d\sigma - \int_A \left( \overline{F^{SH}_{z,surf}} + \overline{F^{LH}_{z,surf}} \right) \left( \frac{1}{T_S} - \frac{1}{T_E} \right) d\sigma \quad , \tag{13}$$

basically because we are overestimating the temperature difference across which the irreversible heat transport takes place. See Appendix A for a detailed derivation of inequality (13). Note that, as said, vertical fluxes are positive upwards. Since at surface energy balance applies, we have that $\overline{F^{SW}_{z,surf}} + \overline{F^{LW}_{z,surf}} + \overline{F^{SH}_{z,surf}} + \overline{F^{LH}_{z,surf}} = -\vec{\nabla}_H \cdot \overline{\vec{H}_{surf}}$, where $\vec{\nabla}_H \cdot \overline{\vec{H}_{surf}}$ is the horizontal divergence of the transport performed by the ocean. By comparing Eqs. (12) and (13), we obtain that the material entropy production by dissipation of kinetic energy is bounded from below by the entropy produced by the large scale horizontal transport of heat:

$$\int_A \frac{\overline{\varepsilon^2}}{T_{diss}} d\sigma > -\int_A \frac{\vec{\nabla}_H \cdot \overline{\vec{H}}}{T_E} d\sigma - \int_A \vec{\nabla}_H \cdot \overline{\vec{H}_{surf}} \left( \frac{1}{T_S} - \frac{1}{T_E} \right) d\sigma \approx -\int_A \frac{\vec{\nabla}_H \cdot \overline{\vec{H}}}{T_E} d\sigma. \tag{14}$$

Under the reasonable hypothesis that $(1/T_S - 1/T_E)$ is approximately constant (roughly corresponding to a spatially homogeneous atmospheric lapse rate), the second term in the first inequality is negligible, since the area integral of the horizontal convergence of subsurface enthalpy flux vanishes. Even with a more conservative scale analysis, we basically obtain the same result, because $|1/T_E| \gg |1/T_S - 1/T_E| = |1/T_E| \times |(T_E - T_S)/T_S|$ and in present day climate the ocean enthalpy transport contributes to about 30% of the total enthalpy transport (Lucarini and Ragone 2011). Equation (14) implies that $\overline{\dot{S}_{mat}^{diss}} > \overline{\dot{S}_{mat}^{hor}}$, i.e. the material entropy production by dissipation of kinetic energy is bounded from below by the entropy produced by the horizontal transport of heat performed by the large scale motion.

Moreover, Eq. (14) allows us to derive an approximate inequality providing a constraint on the intensity of the Lorenz energy cycle $\overline{W} = \overline{D} = \int \overline{\varepsilon^2} dV$. As half of the kinetic energy is dissipated mainly at the boundary layer and half in the free atmosphere (Peixoto and Oort 1992), we have $T_E \leq T_{diss} \leq T_{BL} \approx T_S$. Since $\overline{\varepsilon^2}$ is positive definite and the fractional spatial variation of $T_S$ is relatively small, we derive a lower bound $\overline{W}_{min}$ for the intensity of the Lorenz Energy cycle:

$$\overline{W} = \overline{D} = \int_A \overline{\varepsilon^2} d\sigma \geq -\langle T_S \rangle \int_A \frac{\vec{\nabla}_H \cdot \vec{H}}{T_E} d\sigma \geq -\langle T_E \rangle \int_A \frac{\vec{\nabla}_H \cdot \vec{H}}{T_E} d\sigma = \langle T_E \rangle \overline{\dot{S}_{mat}^{hor}} = \overline{W}_{min} \quad (15)$$

where the square brackets indicate spatial averaging, the second inequality results from $\langle T_S \rangle \geq \langle T_E \rangle$, and we consider the definition of $\overline{\dot{S}_{mat}^{hor}}$ given in Eq. (12). By using $\overline{F_{z,TOA}^{SW}} + \overline{F_{z,TOA}^{LW}} = -\overline{R} = -\vec{\nabla}_H \cdot \vec{H}$, we then express the constraint on the intensity of the Lorenz energy cycle and on the corresponding rate of dissipation of the kinetic energy purely on quantities that can be derived from measurements (or model data) evaluated at the top of the atmosphere as follows:

$$\overline{W}_{min} = -\langle T_E \rangle \int_A \frac{\overline{R}}{T_E} d\sigma. \quad (16)$$

Note that we have defined the TOA radiation balance $\overline{R}$ as positive (negative) where the system behaves as a net absorber (emitter) of radiation, which is tipically the case at low (high) latitudes. It is possible to verify the physical consistency of Eq. (16) considering the following extreme scenario. If the planet has no atmosphere, so that $\overline{W} = 0$, the right term of the inequality must also be vanishing. This is consistent with the fact that in the absence of a fluid envelope, no enthalpy can be transported horizontally, so that when long term averages area considered, $-\overline{R} = -\vec{\nabla}_H \cdot \overline{\vec{H}} = \overline{F_{z,TOA}^{SW}} + \overline{F_{z,TOA}^{LW}} = 0$. This implies that the SW and LW fluxes at the top of the (infinitesimal) atmosphere – as well as at surface - have to be everywhere equal in magnitude and opposite in sign.

*3.1 Baroclinic Efficiency*

We can rewrite Eq. (16) as follows:

$$\overline{W_{min}} = -\langle T_E \rangle \left( \int_{A_>} \frac{\overline{R}}{T_E} d\sigma + \int_{A_<} \frac{\overline{R}}{T_E} d\sigma \right), \tag{17}$$

where we have divided the domain in two regions $A_>$ and $A_<$, the former (latter) describing the subdomain featuring a positive (negative) radiation budget at the top of the atmosphere. We can express Eq. (17) as:

$$\overline{W_{min}} = -\langle T_E \rangle \left( \frac{\langle \overline{R} \rangle_> |A_>|}{T_E^>} + \frac{\langle \overline{R} \rangle_< |A_<|}{T_E^<} \right), \tag{18}$$

where $\langle R \rangle_>$ is the spatial average of the net radiative budget performed on the 2D domain $A_>$ (having measure $|A_>|$), with equivalent notation applying for the negative radiative balance case. Note that $|A_>|+|A_<|=|A|$. Since $\int_A \overline{R}d\sigma=0$, so we have that $\langle \overline{R}\rangle_> |A_>|+\langle \overline{R}\rangle_< |A_<|=0$. Instead, $T_E^>$ and $T_E^<$ are reference temperatures obtained by averaging the emission temperature over the domains $A_>$ and $A_<$, respectively, and using the value of the net radiative budget as weighting function:

$$T_E^{>(<)} = \frac{\int_{A_{>(<)}} \overline{R}d\sigma}{\int_{A_{>(<)}} \frac{\overline{R}}{T_E}d\sigma}. \qquad (19)$$

We then obtain:

$$\overline{W_{\min}} = \langle \overline{R}\rangle_> |A_>|\langle T_E\rangle\left(\frac{1}{T_E^<}-\frac{1}{T_E^>}\right) = \eta_h \langle \overline{R}\rangle_> |A_>|, \qquad (20)$$

with

$$\eta_h = \langle T_E\rangle\left(\frac{1}{T_E^<}-\frac{1}{T_E^>}\right) \approx \frac{T_E^> - T_E^<}{\langle T_E\rangle} \approx \frac{T_E^> - T_E^<}{T_E^>} \qquad (21)$$

where we have assumed that $\langle T_E\rangle \approx 1/2(T_E^> + T_E^<)$ and that $(T_E^> - T_E^<)/\langle T_E\rangle \ll 1$. Whereas the first assumption is quite obvious, since we are averaging over two regions $A_>$ and $A_<$ of analogous size, we underline that the second assumption is usually verified even in the presence of large spatial variability of the radiative balance, as a fourth root is involved in the definition of the emission temperature. The quantity $\eta_h$ links the input of radiative energy at an average rate $\langle \overline{R}\rangle_> |A_>|$ into the warm subdomain at temperature $T_E^>$ to the lower bound to average rate of production of mechanical work, and can be interpreted as the climate Carnot-like efficiency related *only* to differential heating at the top of the atmosphere.

The overall energy balance of the climate system imposes that the fluid envelope of the planet transports through large scale motions an amount of enthalpy $\overline{F} = \langle \overline{R} \rangle_> |A_>|$ from the regions of the climate system featuring a positive radiative budget at the top of the atmosphere to those which constantly lose energy to space (Lorenz 1967, Stone 1978b, Peixoto and Oort 1992, Lucarini and Ragone 2011). Due to the fairly zonal nature of the net TOA radiative balance (Peixoto and Oort 1992), such compensation translates into the fact that in each hemisphere the location of the peak of the meridional enthalpy transport coincides with the latitudinal boundary dividing the radiatively heated low latitudes and the radiatively cooled high latitudes in the northern (southern) hemisphere. Moreover, since the two hemispheres are rather similar in terms of average zonal energy budgets and inferred meridional transports, as imposed by the constraints given in Stone (1978) and confirmed by Lucarini and Ragone (2011), the intensity of the peak of the transport in either hemisphere is approximately $1/2 \langle \overline{R} \rangle_> |A_>|$. The lower bound to the intensity of the Lorenz energy cycle given in Eq. (20) can be interpreted as product of an efficiency related to meridional temperature differences and the fluxes across such meridional gradients. This provides further theoretical insight and suitable conceptual framework to the intuition by Barry et al. (2002) on the possibility of defining a "baroclinic heat engine" extracting work from the meridional heat flux.

What presented here gives the basic ingredients for providing a rigorous construction of the simplified two-box model usually considered in the literature (Lorenz et al. 2001). Such a reduced model, which allows for the description of entropy production due to horizontal processes of heat exchange only, is enclosed in a dashed rectangle in Fig. 1. The warm box (Box 1) is defined by the portion of the fluid envelope of the climate system featuring a positive net radiative balance at TOA and has a temperature equal to $T_E^>$. The cold box (Box 2) is defined by the part of the fluid envelope of the planet featuring a negative net radiative balance at TOA, i.e. the mid-high latitudes, and has temperature equal to $T_E^<$. The irreversible heat transfer $\overline{F} = \langle R \rangle_> |A_>|$ from warm to cold areas generates entropy at rate:

$$\overline{\dot{S}_{mat}^{hor}} = \langle \overline{R} \rangle_> |A_>| (1/T_E^< - 1/T_E^>) = \eta_h \overline{F}/\langle T_E \rangle \tag{22}$$

*3.2 Bound on the degree of irreversibility of the system*

The bound $\overline{\dot{S}_{mat}^{diss}} > \overline{\dot{S}_{mat}^{hor}}$ obtained in Eq. (14) can be used to introduce a further bound to the thermodynamical properties of the system. With a trivial manipulation of the expression of the parameter of irreversibility $\alpha$, we obtain:

$$\alpha = \frac{\overline{\dot{S}_{mat}} - \overline{\dot{S}_{mat}^{diss}}}{\overline{\dot{S}_{mat}^{diss}}} < \frac{\overline{\dot{S}_{mat}} - \overline{\dot{S}_{mat}^{hor}}}{\overline{\dot{S}_{mat}^{diss}}} \approx \frac{\overline{\dot{S}_{mat}^{vert}}}{\overline{\dot{S}_{mat}^{diss}}} = \alpha_{max} \tag{23}$$

where $\alpha_{max}$ is the upper bound to the parameter of irreversibility. Observing that the Bejan number *Be*, a widely used parameter in the applied thermodynamics literature (Paoletti et al. 1989), can be written as $Be = \alpha + 1$, we obtain:

$$\frac{\overline{\dot{S}_{mat}}}{\overline{\dot{S}_{mat}^{hor}}} \approx \frac{\alpha_{max} + 1}{\alpha + 1} = \frac{Be_{max}}{Be} \approx \frac{\overline{\overline{W}}}{\overline{\overline{W}}_{min}} \tag{24}$$

so that the product of the lower bound to the intensity of the Lorenz energy cycle and of the upper bound of the Bejan number is equal to the product of the actual Bejan number times the actual intensity of the Lorenz energy cycle.

## 4. Estimation of the Thermodynamic Properties of PCMDI/CMIP3 Climate Models

The theory developed in the previous sections is applied to analyse the thermodynamic properties of state-of-the-art Climate Models (CMs) using the publicly available output from the PCMDI/CMIP3 dataset (http://www-pcmdi.llnl.gov/). We have used 100 years long time series of monthly means of surface temperature and of radiative fluxes (long-wave and short-wave) at the surface and at the top of the atmosphere, from the PI control run scenario. Since for one of these models we have an exact computation of the material entropy production budget in the PI scenario (Pascale et al. 2009), we can also test the accuracy of our approximations in these conditions. The PCMDI/CMIP3 dataset includes data from over 20 CMs, but only 14 CMs were considered in this analysis, due to lacking of some fields and/or inconsistencies in the dataset. Models making use of flux adjustments have been excluded too, since they provide an unphysical representation of the thermodynamics of the climate system. See table 3 for the list of models which have been used in this paper. Each model is labeled with a number as in Lucarini and Ragone (2011).

In PI conditions, all the parameters of the model are kept constant (in particular the $CO_2$ concentration is fixed at 280 ppm), so that the slowest forcing acting on the system is given by the seasonal cycle (the solar cycle, albeit very weak, is considered in some models). Therefore, integrating the model over a sufficiently long time period, the system should reach a stationary state. Once the steady state is reached, the minimal time averaging window over which one can expect time-independent statistical properties is given by one year. In the following, the time-averaging operator $\overline{(\bullet)}$ used throughout the formulas derived in this paper is considered to act over one year. The consideration of one hundred years for each CM allows constructing a suitable, robust statistics for the yearly-averaged thermodynamical properties of the corresponding climate.

In general, the stationary state of a non-equilibrium system is characterized by vanishing global balances of energy and entropy (Lucarini 2009) and by time-independent statistical properties for the state variables of the system. While this second condition is fulfilled by the climate models here considered, basically by definition of PI control run, the first condition is, in general, not satisfied, as investigated in Lucarini and Ragone (2011), where significant biases have been found in the global energy balances of the system and of its principal subsystems for these models in PI conditions. Nevertheless, as discussed below, it is still possible to take care of such unphysical biases in a similar way as done in Lucarini and Ragone (2011) when computing the meridional enthalpy transport.

There are algebraically different ways to define the yearly value of the emission temperature field $T_E$. Since the multiplicative factor to the annual mean of the radiative balance in Eqs. (11) and (12) is $1/T$, we have computed the monthly emission temperatures $T_E^m = \sqrt[4]{F_{z,TOA}^{LW,m}/\sigma}$, where $F_{z,TOA}^{LW,m}$ is the monthly mean of the longwave emission at TOA, and than we have computed $T_E$ as the inverse of the annual mean of the inverse of the $T_E^m$

$$T_E = \left(\overline{1/T_E^m}\right)^{-1} \qquad (26)$$

Rather than considering the actual surface temperature fields provided by the CMs, in order to be consistent with the idea of estimating the thermodynamics properties starting from radiative fields only, the considered surface temperature field $T_S$ has been computed from the outgoing longwave radiation at surface by mirroring the procedure described above for computing the 2D $T_E$ fields starting from the TOA outgoing longwave radiation, thus assuming an unit value for emissivity everywhere and at all times. We have verified that the consideration of the surface temperature fields given as outputs by the CMs impacts our results in an entirely negligible way.

The presence of a spurious bias in the TOA global energy balance has been automatically cured subtracting at each grid-point in each year the globally averaged TOA global energy imbalance. This is a standard procedure when inferring the meridional enthalpy transports, as discussed in Carissimo et al. (1985) and Lucarini and Ragone (2011), and it is justified by the fact that the global energy imbalance at each year is small if compared to the latitudinal variability of the energy balance (of the order of 1 $Wm^{-2}$ vs. 100 $Wm^{-2}$), which is the quantity we are mostly interested into. When considering entropy estimators, it is especially important to remove the bias in the TOA energy balance, because the related relative error on the estimate of the horizontal component of the material entropy production would be large.

In the case of the net radiative balance at the surface we do not have a zero-sum constraint (latent and sensible heat are not involved in our calculations) because, on the contrary, the typical value of the net global radiative balance at the surface is of the order of 100 $Wm^{-2}$. Thus, it is not possible to define a bias by observing the violation of the zero mean constraint. In any case, even if the bias were of the order of 10 $Wm^{-2}$, the resulting error in estimating the vertical contribution to the material entropy production would be of the order of 5%, so that we can safely ignore it.

We have then computed for each model the yearly value of the total material entropy production $\overline{\dot{S}_{mat}}$, of its vertical and horizontal components $\overline{\dot{S}_{mat}^{ver}}$ and $\overline{\dot{S}_{mat}^{hor}}$, of the equivalent temperatures $T_E^<$ and $T_E^>$, of the baroclinic efficiency $\eta_h$, and of the lower bound to the intensity of the Lorenz energy cycle $\overline{W_{min}}$. From the obtained 100-year time series we have then estimated the expectation value and the confidence interval of the mean of each quantity using the block-bootstrap resampling technique (see, e.g., Lucarini and Ragone 2011). For all the considered time series the standard deviation is very small, so that the width of the 95% confidence interval is in all cases below 1% of the expectation value. In Table 1 we provide estimates of the expectation values of the considered thermodynamic parameters.

*4.1 Mean values*

The entropy budget of CM 13 has been extensively analysed in PI conditions in Pascale et al. (2009), so that we have an excellent test for the accuracy of our approximated formulas. Formula (13) gives an estimate for the global material entropy production for CM 13 of 54.0 $mWm^{-2}K^{-1}$, which is in excellent agreement with the correct value of 51.8 $mWm^{-2}K^{-1}$ obtained by Pascale et al. 2009. This single test is highly encouraging in suggesting that our approximate approach provides an accurate guidance and rather stringent estimates of the actual material entropy production of the system, with an uncertainty of about 5%.

For explanatory purposes, in Fig. 2a-b we present maps of the average rate of material entropy production by vertical and horizontal processes, respectively, while in Fig. 2c we show the field of emission temperature $T_E$. All outputs are obtained from CM 13, even if all CMs give similar pictures.

In Fig 2a we observe that, consistently with the discussion in Section 2.3, the spatial field whose integral gives $\overline{\dot{S}_{mat}^{ver}}$ is positive everywhere except in small areas at high elevation and latitude where very small negative contributions are obtained. These unphysical results are due to the large temperature inversion observed in these areas, which somewhat compromises the scaling analysis we have adopted. In any case, the global effect of these contributions is entirely negligible. As expected, high values are observed where intense evaporation is present, as in the warm pool of the western Pacific and Indian Ocean, whereas, consistently, very low values are observed in the cold tongue of the Eastern Pacific, near western boundary currents, and in the temperate and cold oceans, while the Mediterranean Sea stands out as a warm pool. Land areas typically feature much lower values than ocean areas at similar latitudes, except for areas characterized by warm and moist climate, such as in the equatorial forests of Amazon, Congo, South-Eastern Asia, where water is always available for evaporation and ocean-like values are obtained. The relevance of the vertical latent heat transport in determining the entropy production is also clarified by the fact that values close to zero are found in deserts (Sahara, Kalahari, Central Asia, Southwestern US and Mexico, Southern Australia, Patagonia), even if intense sensible heat exchanges take place, and in mid-high latitudes terrestrial areas. At polar latitudes, vanishing values are obtained since convective processes are very weak and moisture is almost absent.

Figure 2b shows that the globally positive value of material entropy production due to horizontal processes results from the (almost perfect, see below) compensation between positive and negative values, with the former dominating the mid-high latitudes and the latter present in the equatorial and tropical regions. Interestingly, some of the features observed in Fig. 2a are found also here: the areas of vigorous moist convection appear as areas of strong negative values related to divergence of (mostly) latent heat. It is also important to note that deserts feature positive values, as they cannot contribute to the latent heat transport and are characterized by high albedo.

Figure 2c shows that, generally, lower (higher) emission temperature $T_E$ are found in the $A_<$ ($A_>$) area, which clarifies that $T_E^> > T_E^<$ (see below). Nonetheless, local violations to the simple rule "$A_> \rightarrow$hot & $A_< \rightarrow$cold" are found. In fact, the Intertropical Convergence Zone (included in $A_>$) features relatively low emission temperatures, since deep convection creates LW radiation-opaque clouds at very high altitude, whereas, conversely, over deserts (included in $A_<$) very high emission temperatures are found, because of the low cloud coverage.

In Fig. 3 we present a scatter plot of the globally averaged, annual mean values of its vertical and horizontal component superimposed on the isolines of the total material entropy production. The error bars represent the 95% confidence interval of the estimate. Models are labeled with numbers as in Table 3. We can see that, apart from model 10, the typical value of the annual material entropy production is between 52 and 58 $mWm^{-2}K^{-1}$. These figures match well with the approximate estimate by Ambaum (2010). The contributions to material entropy production due to vertical and horizontal processes typically amount to about 50 $mWm^{-2}K^{-1}$ and about 5 $mWm^{-2}K^{-1}$, respectively, so that the contribution due to vertical processes is dominant by about one order of magnitude.

Figure 3 suggests that the anomalously high total material entropy production of CM 10 is due to a large contribution due to vertical processes, while the contribution to horizontal component is consistent with the values of the other models. In Lucarini and Ragone (2011), CM 10 was found to be the only model with a negative annual global oceanic energy balance. This suggests that an excess of energy is transported into the atmosphere, mostly due to convective processes, with a resulting positive anomaly in entropy production. This hints at some problems in the representation of convective processes, whose parameterisation is especially problematic in a coarse resolution model such as CM 10. Quite reassuringly, the figure we obtain for the total material entropy production for CM 10 agrees with what found by Goody (2000) in a similar version of the same model.

CM 6 features a small horizontal component of the material entropy production compared to the other models, even if this does not impact substantially the value of the total material entropy production, given the small weight of the contribution of the horizontal processes. In Lucarini and Ragone (2011), CM 6 was found to have the position of the peak of the annual meridional enthalpy transport located anomalously near the equator with respect to the other models (most notably in the northern hemisphere). This is consistent with CM 6 having a small material entropy production from horizontal processes, since in this model most of the transport is realized at lower latitudes, where the meridional temperature gradient is smaller.

In Fig. 4 we present a scatter plot of $T_E^<$ and $T_E^>$ together with the corresponding isolines of the efficiency $\eta_h$, which are almost indistinguishable from straight lines. We can see that models feature values of $T_E^<$ between around 240 and 244 K, and values of $T_E^>$ between around 255 K and 260 K, so that the typical equivalent temperature difference is about 15 K. Lucarini and Ragone (2011) proposed that spurious positive energy imbalances at steady state due to inconsistencies in the treatment of energy exchanges throughout the climate system induce the presence of a cold bias when emission temperatures are considered. In agreement with this, in the present analysis we find that colder CMs are for the most part those featuring rather large positive global energy imbalances in Lucarini and Ragone (2011), whereas the warmer CMs feature average global energy balances close to zero.

Most models feature values of the efficiency $\eta_h$ between 0.050 and 0.060, with few models in the range 0.040-0.050 on one side and 0.060-0.065 on the other side. Since CMs differ much more on $T_E^>$ than on $T_E^<$, the largest and smallest efficiencies are related basically to very high and very low values of $T_E^>$, respectively, so that the low latitudes seem to have a prominent role in determining the efficiency of the baroclinic engine. By comparing the figures reported in Table 1 on the value of $\overline{F} = \langle \overline{R} \rangle_> |A_>|$ and the sum of the peaks of the meridional transports in the northern and southern hemispheres given in Lucarini and Ragone (2011), we find that the meridional heat transport provides for all CMs a contribution of about 95% to the total transport from radiatively warmer to radiatively cooled areas of the climate system (the remaining 5% is due to zonal heat fluxes). Therefore, the interpretation of $\eta_h$ as the baroclinic efficiency introduced by Barry et al. (2002) is definitely appropriate.

In Table 1 we also provide for all CMs the estimates of the lower bound to the intensity of the Lorenz energy cycle for all considered CMs, computed according to Eq. (16). Most values span the range 1.0-1.5 Wm$^{-2}$, which definitely captures the right order of magnitude of the Lorenz energy cycle (Peixoto and Oort 1992). Fortunately, the scientific literature provides some benchmarks to be used to test how stringent our bounds are. In the case of CM 13, Pascale et al. (2009) report that the intensity of the Lorenz energy cycle is about 3.1 Wm$^{-2}$ in simulations mirroring exactly the PI conditions simulation considered here. This implies that the lower bound underestimates the actual value by 60%. In the case of CM 4, CM 18 and CM 20, Marquez et al. (2010) provide estimates for the intensity of the Lorenz energy cycle of 2.7, 3.1 and 3.3 Wm$^{-2}$, respectively, even if the data are referred to XX century simulations rather than PI conditions. Nevertheless, since relatively small changes of CO2 concentration seem to impact only marginally the Lorenz energy cycle (Lucarini et al. 2010b, Hernandez-Deckers and von Storch 2010), we conclude that also for these models the lower bound underestimates the actual value by around 60%. These results suggest that our approach is fundamentally correct and the theoretically obtained lower bound provides a good zero-order approximation of the actual value of the intensity of the Lorenz energy cycle.

*4.2 Variability*

We now analyse the mutual correlations of the time series of the yearly values of some of the thermodynamic parameters we have derived, in order to improve our understanding of the dynamical processes keeping the system at a well-defined stationary state. Our time series are 100 years long and feature very weak memory, so that in all cases the 95% confidence interval on the estimates of the correlations has a half-width of 0.2.

First, we look at potential feedbacks of the system. In Fig. 5 we present a scatter plot of the correlation between the yearly time series of the baroclinic efficiency of the system $\eta_h$ and of the total large scale horizontal transport $\overline{F} = \langle \overline{R} \rangle_{>} A_{>}$ versus the correlation of the yearly time series of the horizontal and vertical contributions to the material entropy production. The uncertainty range of the null-hypothesis is represented by the cross centered in (0,0). We find that nearly all the models feature a statistically significant positive correlation between the horizontal and vertical components of the material entropy production, even if the range of values is quite wide. This implies that there is not such a thing as compensation between the two components, with positive anomalies of one component typically corresponding to negative anomalies of the other component, thus determining a negative feedback minimizing the variability of the total material entropy production. On the contrary, it is apparent that during the years where the total material entropy production has, e.g. a positive anomaly, both components change accordingly. Fig. 2a-b, is consistent with the idea that the strength of the large scale tropical convection (which mostly contributes to $\overline{\dot{S}_{ver}^{mat}}$) and of the extratropical dynamics (which mostly contributes to $\overline{\dot{S}_{hor}^{mat}}$) are positively correlated. The positive covariance of the two components of the material entropy production implies that their ratio has a small variability, which, following the discussion presented in subsection 3.2, suggests that the degree of irreversibility α is a well-constrained parameters of the system. The only qualitative exception is CM 10, which features a borderline statistically significant negative correlation between the vertical and the horizontal components. Combing this result with the evidence given above on the anomalously high value of material entropy production due to vertical processes (see Fig. 3), we propose that CM10 might present some issues in the treatment of vertical exchange processes (basically convection) and in their coupling with large scale processes responsible for horizontal transport.

Looking at the x-axis in Fig. 5, we discover that among CMs the values the correlation between $\overline{F}$ and $\eta_h$ span a rather wide range, with the majority of models featuring a statistically significant negative correlation. Since the efficiency is a normalized measure of the equivalent temperature difference between the warm and the cold box, it is also an integrated measure of the effective meridional temperature gradient realized at the stationary state. Therefore, the presence of a negative correlation between the efficiency and the meridional total transport suggests that a negative feedback acts to dampen large fluctuations of the meridional temperature gradient, in broad agreement with the baroclinic adjustment theory by Stone (1978b).

Nevertheless, we observe that for several models the coupling between the two main major features involved in the large scale re-equilibration process - the heat transport and the temperature gradient – is quite weak. These findings seem to be only partially in agreement with the presence of a feedback keeping $\overline{\dot{S}_{mat}^{hor}}$ near an extremum, as implied by the MEPP hypothesis (Kleidon and Lorenz 2005).

Since the value of $\overline{\dot{S}_{mat}^{hor}}$ is proportional to the product of $\overline{F}$ times the temperature difference between the warm and the cold box (see Eq. (19)), and since the correlation between $\overline{F}$ and $\eta_h$ is in general not very strong, it seems of relevance to check out which of the two factors $\overline{F}$ and $\eta_h$ is more strongly correlated to $\overline{\dot{S}_{mat}^{hor}}$. This would help answering the question on whether the variability of the entropy production is driven more by the temperature differences or by the total heat flux.

In Fig. 6 we see that all CMs feature a very strong positive correlation between the horizontal component of the material entropy production and the efficiency. All values are above 0.7 and several CMs feature correlations above 0.9. Instead, the correlation with the total transport spans over a larger set of values (from -0.5 to 0.6), so that qualitatively different properties are found among CMs. Surprisingly, only a minority of CMs feature a statistically significant positive correlation between $\overline{\dot{S}_{mat}^{hor}}$ and $\overline{F}$. Therefore, the presence of an anomalously high efficiency or, equivalently, of a high equator-to-pole temperature gradient is definitely a better statistical predictor of high entropy production due to horizontal processes. This result suggests that simplified parameterisations of entropy production could be written efficiently in the form $\overline{\dot{S}_{mat}^{hor}} = \overline{\dot{S}_{mat}^{hor}}(\eta_h)$, which provides further evidence of the relevance of the efficiency parameter introduced here.

These results definitely suggest that, as opposed to the climatological averages considered above, CMs are much less consistent with each other in the representation of the second moments of the large scale thermodynamical properties. This hints at the need for further explorations of the related climate feedbacks.

## 5. Conclusions

In this paper we have performed a re-examination of material entropy production in the climate system by proposing theoretical advances and by presenting new results derived from control runs of state-of-the-art CMs.

We have first discussed various approaches recently discussed in the literature for analyzing the entropy budget in the climate system and concluded, from the critical appraisal of recent results presented by Pascale et al. (2009), that the approach of considering a simplified description of a "moist" atmosphere, where water is treated mainly as a passive substance which provides/removes latent heat, is well suitable when studying the material entropy production.

We have introduced an approximate splitting between material entropy production contributions due to vertical processes, mostly related to convection and characterized by short time scales, and those due eminently to horizontal processes, mostly related to the large scale motions in the climate system and characterized by longer time scales. Notably, such an approach allows for computing with good precision the material entropy production of the climate system using only 2D radiative fields at the top of the atmosphere and at surface (even if the evaluation of the latter can be experimentally challenging). By comparing our approximate formulas results with the "quasi-exact" treatment by Pascale et al. 2009, we find that our 2D, simplified calculations are correct to within 5%.

We have derived bounds to basic thermodynamical properties of the climate system based upon time averaged TOA radiative data only. This could in principle be particularly promising especially when one deals with planetary objects, such as distant extra-solar planets, where the amount of observational data is limited. We have proved that the material entropy produced by the horizontal large scale enthalpy transports is a lower bound to the total material entropy production and, in particular, to its portion related to the dissipation of mechanical energy. From this, a lower bound is derived for the average rate of the Lorenz energy cycle. The Lorenz energy cycle results to be larger than the product of the net input of radiative energy in the positive energy balance regions times a suitably defined baroclinic efficiency, proportional to the difference between the average emission temperatures of areas of the planet with positive and negative radiative balance at the top of the atmosphere. This provides a generalization of the idea that the atmospheric circulation can be described as resulting from a baroclinic heat engine extracting work from the meridional heat flux (Barry et al. 2002). As the meridional heat flux by atmospheric eddies contributes substantially to the total meridional heat flux and, at the same time, plays a fundamental role in the Lorenz Energy cycle by transforming zonal into eddy available potential energy, the presence of such constraints is not so surprising.

We have clarified the relevance of 2-box models representative of planetary circulations discussed in, e.g., Lorenz et al. (2001). These models, by neglecting vertical processes (which, in the case of our planet, give the dominant contribution to the material entropy production), cannot provide a plausible description of the material entropy production of the climate system, nor can be used to test MEPP. Instead, a minimal model of material entropy production in a planetary system (schematically depicted in Fig. 1) requires at least four boxes, representing cold and warm pools of fluid and cold and warm surfaces beneath, and taking care of representing both horizontal and vertical transport processes.

The evaluation of the contributions to the material entropy production due to horizontal and vertical processes on state-of-the-art CMs runs provides rather interesting insights. We discover that models agree within about 10% on the value of total material entropy production, and specifically, the disagreements are within about 10% on the value of the vertical term, which is the largely dominant one, whereas larger disagreements (of the order of 20%) exist on the horizontal term, which is about one order of magnitude smaller. Two models seem to be somewhat out of this picture.

As the entropy production due to horizontal processes can be interpreted as resulting from the large scale heat transport from low (warm) to high latitude (cold) regions, its intensity depends at first order on the product between the maximum intensity of the meridional transport times the differences between the emission temperature of the two regions. Obvious physical balances would suggest that the two factors should be, model-wise, negatively correlated, so that one expects that the agreement among CMs should be better for the entropy production rather than for the transport diagnostics. Instead, in Lucarini and Ragone (2011) it is shown that discrepancies on the total (and atmospheric) meridional transports among models are also around 20% - it could be argued a role is played by disagreements on their meridional gradients of albedo (see Probst et al. 2010 for a related analysis on total cloud cover). As a result, the baroclinic efficiency of the models disagrees substantially, as do the lower bounds to the intensity of the Lorenz energy cycle. For the CMs where independent data on the intensity if the Lorenz energy cycle are available, we find that such bounds are relatively stringent (within a factor of about 2), which supports the relevance of our theoretical results.

We have then explored the second moments of the thermodynamic parameters discussed above by looking at the properties of their correlations, in order to grasp some understanding of the large feedbacks acting on the system. We discover that CMs disagree quantitatively to a much greater extent than when looking at first moments of the considered of the thermodynamic parameters. Nevertheless, a robust feature of most models is the positive correlation of the contributions to material entropy production due to vertical and to horizontal processes. The dynamical link between the tropical and the extratropical dynamics suggests that no compensation mechanism is in place to *control* the total material entropy production. Similarly, most models feature a negative correlation between the intensity of the large scale heat transport and the difference between the effective emission temperatures of the radiatively heated and cooled areas of the planet, in broad agreement with the theory of baroclinic adjustment. Finally, we discover that such temperature difference (or, equivalently, the baroclinic efficiency of the system) is much stronger proxy for the material entropy production due to horizontal processes than the heat flux, thus suggesting its great relevance of large scale indicator of the climate system.

Future investigations will move along the following lines. First, it is of great interest to study how the increase of greenhouse gases impacts the details of the material entropy production on Earth by taking advantage of the PCMDI/CMIP3 dataset. While we expect an overall positive sensitivity of such thermodynamic parameter - see Lucarini et al. (2010b) – it will be interesting to analyse how climate change effects the contributions to material entropy production due to vertical and horizontal processes. Preliminary analyses on the SRESA1B scenario runs suggest that the impact has opposite sign for the two terms, with an increase in entropy production due to vertical processes occurring in all CM as a result of enhanced convection. This result, which is apparently at odds with the positive covariability of the annual time series described above in the PI run, casts further doubts in the straightforward applicability of the properties of the natural variability of the climate system properties for inferring the properties of long term, forced climate variations (Lucarini 2008b,2009b). As suggested by the maps showing the spatial properties of the thermodynamic fields, the theory presented in this paper can be used also for devising diagnostic tools to be used to analyze local processes and specific geographical features of the climate, both using satellite data and numerical model data. Nonetheless, this requires re-examining in more quantitative terms the scale analysis discussed in this paper and assess its limitations,

Second, a detailed examination of the Lorenz energy cycle and entropy production of other celestial objects of the solar system seems of great relevance for its own sake and for understanding the relevance of the thermodynamical bounds proposed here. It is encouraging to note that various models belonging to the PLASIM family (Fraedrich et al. 2005) have already been adapted to study the atmospheres of Titan (Grieger et al. 2004) and Mars (Stenzel at al. 2007), thus allowing for an integration of satellite and model data. Moreover, observations are providing a growing amount of information on extrasolar planets, whose investigation provides an exciting scientific challenge. Obviously, as the chemical composition, the characteristic temperatures of the atmospheres of planets and their astronomical parameters present a great variety, the scaling analyses presented here should be carefully reconsidered.

Third, it would be important to reconcile the definition of thermodynamic efficiency proposed by Johnson (2000) and Lucarini (2009a) with that proposed by Ambaum (2010). In this direction, concepts borrowed from endoreversible thermodynamics (Hoffman et al. 1997) could be of great help.

## Acknowledgements

The authors wish to thank S. Pascale for insightful comments on an earlier version of the paper and the reviewers for constructive criticism which has greatly improved the quality of the presented results. VL wishes to acknowledge the support of the FP7-ERC Grant NAMASTE "Thermodynamics of the Climate System".

**APPENDIX**

In the following we derive inequality (13), which has served as a starting point for obtaining the lower bounds to the entropy production due to the dissipative processes and to the intensity of the Lorenz energy cycle. In section 2 we have shown that the global material entropy production of the Earth system at the steady state can be written as

$$\overline{\dot{S}_{mat}} = \int_A \frac{\langle \overline{\varepsilon^2} \rangle}{T_{diss}} d\sigma - \int_A \overline{F^{SH}_{z,surf}} \left( \frac{1}{T^+_{SH}} - \frac{1}{T^-_{SH}} \right) d\sigma - \int_A \overline{F^{LH}_{z,surf}} \left( \frac{1}{T^+_{LH}} - \frac{1}{T^-_{LH}} \right) d\sigma + \Delta \overline{\dot{S}^{LH}_{mat}} \qquad \text{A.1}$$

where we recall that $\langle \overline{\varepsilon^2} \rangle$ is the vertically integrated dissipational heating (positive defined), $\overline{F^{SH}_{z,surf}}$ and $\overline{F^{LH}_{z,surf}}$ are the vertical components of the sensible and latent heat fluxes at the surface (positive upwards), $T_{diss}$ is the characteristic temperature at which the dissipational heating occurs, $T^+_{SH}$ and $T^+_{LH}$ are the characteristic temperatures at which sensible and latent heat are extracted, $T^-_{SH}$ and $T^-_{LH}$ are the characteristic temperature at which sensible and latent heat are absorbed, and $\Delta \overline{\dot{S}^{LH}_{mat}}$ represents the correction related to the horizontal transport of water vapor. Defining

$$\Delta \overline{\dot{S}^{corr}_{mat}} = -\int_A \overline{F^{SH}_{z,surf}} \left( \frac{1}{T^+_{SH}} - \frac{1}{T^-_{SH}} + \frac{1}{T_E} - \frac{1}{T_S} \right) d\sigma - \int_A \overline{F^{LH}_{z,surf}} \left( \frac{1}{T^+_{LH}} - \frac{1}{T^-_{LH}} + \frac{1}{T_E} - \frac{1}{T_S} \right) d\sigma \qquad \text{A.2}$$

we can rewrite equation A.1 as

$$\overline{\dot{S}_{mat}} = \int_A \frac{\langle \overline{\varepsilon^2} \rangle}{T_{diss}} d\sigma - \int_A \left( \overline{F^{SH}_{z,surf}} + \overline{F^{LH}_{z,surf}} \right) \left( \frac{1}{T_S} - \frac{1}{T_E} \right) d\sigma + \Delta \overline{\dot{S}^{corr}_{mat}} + \Delta \overline{\dot{S}^{LH}_{mat}} \qquad \text{A.3}$$

Performing a scale analysis we can compare the relative magnitude and the sign of $\Delta \overline{\dot{S}^{corr}_{mat}}$ and $\Delta \overline{\dot{S}^{LH}_{mat}}$ in order to understand which is the relation between $\overline{\dot{S}_{mat}}$ and the first two terms on the r.h.s. of equation A.3.

In order to estimate $\Delta \overline{\dot{S}_{mat}^{corr}}$ we make the following assumptions:

- both sensible and latent heat are extracted at surface temperature $T_{SH}^+ = T_{LH}^+ = T_S$

- the turbulent transport of sensible heat is basically a local process, taking place moslty in the boundary layer, so that $T_{SH}^- = T_{BL} > T_E$

- the water vapour saturation mixing ratio strongly increases with temperature, and on Earth the atmospheric temperature decreases with height. Therefore the vertical scale of water vapour in globally saturated conditions is smaller than that of the atmosphere, and we have that the latent heat is absorbed by the atmosphere at a condensation temperature $T_{LH}^+ = T_C > T_E$

Thus we have

$$\Delta \overline{\dot{S}_{mat}^{corr}} = -\int_A \overline{F_{z,surf}^{SH}} \left( \frac{1}{T_E} - \frac{1}{T_{BL}} \right) d\sigma - \int_A \overline{F_{z,surf}^{LH}} \left( \frac{1}{T_E} - \frac{1}{T_C} \right) d\sigma < 0 \qquad \text{A.4}$$

because both integrals in A.4 are positive (we remember that the vertical fluxes are positive upwards). We can estimate the absolute value of $\Delta \overline{\dot{S}_{mat}^{corr}}$ using the following characteristic values

$T_{BL} \approx 280\ K$, $T_C \approx 260\ K$, $T_E \approx 250\ K$, $\left| \overline{F_{z,surf}^{SH}} \right| \approx 20\ Wm^{-2}$, $\left| \overline{F_{z,surf}^{LH}} \right| \approx 80\ Wm^{-2}$

We obtain

$$\left| \Delta \overline{\dot{S}_{mat}^{corr}} \right| \approx \left| \overline{F_{z,surf}^{SH}} \right| \frac{|T_{BL} - T_E|}{|T_{BL}||T_E|} + \left| \overline{F_{z,surf}^{LH}} \right| \frac{|T_C - T_E|}{|T_C||T_E|} \approx 20 \cdot 10^{-3}\ Wm^{-2} K^{-1} \qquad \text{A.5}$$

In section 2 we have seen that

$$\Delta \overline{\dot{S}_{mat}^{LH}} = -\int_A \overline{F_{out}^{LH}} \left( \frac{1}{T_C} - \frac{1}{T_C'} \right) d\sigma \qquad \text{A.6}$$

where $\overline{F_{out}^{LH}}$ is the amount of latent heat extracted in a certain atmospheric column which is not absorbed in the same column, $T_C$ is the characteristic temperature of absorption of latent heat as above, and $T_C'$ is the effective temperature at which $\overline{F_{out}^{LH}}$ is absorbed (see discussion in section 2). We can give the absolute value of this correction an upper bound as follows:

$$\left|\Delta\overline{\dot{S}_{mat}^{LH}}\right| < \left|\overline{F_{max}^{LH}}\right|\frac{\left|\Delta^{hor}T_C\right|}{\left|T_C\right|^2} < 5\cdot 10^{-3}\ Wm^{-2}K^{-1} \qquad \text{A.7}$$

where $\left|\overline{F_{max}^{LH}}\right| \approx 10\ Wm^{-2}$ is the annual mean of the net transport of latent heat divided by the area of the Earth (Peixoto and Oort 1992), $\left|\Delta^{hor}T_C\right| \approx 30\ K$ is a "generous" measure of the difference of the middle atmospheric temperature between the areas of the planet where the moisture is diverging and converging, and $T_C \approx 260\ K$. We note that in this estimate we have considered that all the moisture which is not condensed in the same column where it is evaporated is condensed at a lower (or higher) condensation temperature. On Earth part of the moisture which evaporates in the tropics is transported and condensed at the equator at a higher condensation temperature (negative contribution to $\Delta\overline{\dot{S}_{mat}^{LH}}$), while part is condensed at the midlatitudes at a lower condensation temperature (positive contribution to $\Delta\overline{\dot{S}_{mat}^{LH}}$), so that the resulting partial compensation ensures us that the obtained bound is even safer. The sign of $\Delta\overline{\dot{S}_{mat}^{LH}}$ depends on which of the two cases gives the dominant contribution.

With this analysis we have that $\Delta\overline{\dot{S}_{mat}^{corr}}$ is negative and in absolute value much larger than the upper bound of $\Delta\overline{\dot{S}_{mat}^{LH}}$. Therefore $\Delta\overline{\dot{S}_{mat}^{corr}} + \Delta\overline{\dot{S}_{mat}^{LH}} < 0$ and we deduce

$$\overline{\dot{S}_{mat}} < \int_A \frac{\left\langle\overline{\varepsilon^2}\right\rangle}{T_{diss}}d\sigma - \int_A \left(\overline{F_{z,surf}^{SH}} + \overline{F_{z,surf}^{LH}}\right)\left(\frac{1}{T_S} - \frac{1}{T_E}\right)d\sigma \qquad \text{A.8}$$

**Figure Captions**

**Figure 1**: Minimal conceptual for the material entropy production of a planetary systems as deduced from Eq. (11). Boxes 1 and 2 represent warm (low latitudes) and cold (high latitudes) fluid domains, coupled by enthalpy transport. Boxes 3 and 4 represent warm (low latitudes) and cold (high latitudes) surface domains, coupled vertically to the corresponding fluid boxes, but not to each other. The dashed rectangle encloses the reduced two-box model usually considered in previous literature.

**Figure 2**: CM 13's time-averaged spatial fields of a) material entropy production via vertical processes $\overline{\dot{S}_{mat}^{vert}}$ (in $WK^{-1}m^{-2}$); b) material entropy production via horizontal processes $\overline{\dot{S}_{mat}^{hor}}$ (in $WK^{-1}m^{-2}$); c) emission temperature (values in $K$). In b) and c) the solid black line separates the area $A_>$ with positive net energy budget at TOA from the area $A_<$ where the net budget is negative.

**Figure 3**: Material Entropy Production in PCMDI/CMIP3 climate models in the pre-industrial scenario runs. The contributions due to horizontal ($\overline{\dot{S}_{mat}^{hor}}$) and vertical ($\overline{\dot{S}_{mat}^{vert}}$) processes are depicted. Isolines of the total value of the material entropy production are shown as solid lines.

**Figure 4**: Equivalent temperatures of the warm ($T_E^>$) and cold ($T_E^<$) boxes of PCMDI/CMIP3 climate models in Pre-Industrial conditions. Isolines of efficiency $\eta_h$ are indicated with solid lines.

**Figure 5**: Strength of the large scale thermodynamic feedbacks for PCMDI/CMIP3 climate models in Pre-Industrial conditions. Correlation between total large scale heat transport $F$ and the baroclinic efficiency $\eta_h$ (x-axis) vs correlation between horizontal ($\overline{\dot{S}_{mat}^{hor}}$) and vertical $\dot{S}_{mat}^{ver}$ components of the material entropy production (y-axis).

**Figure 6**: Correlations between horizontal component of material entropy production ($\overline{\dot{S}_{mat}^{hor}}$) and total meridional transport $\overline{F}$ (x-axis) and efficiency $\eta_h$ (y-axis).

**Table 1:** Thermodynamic parameters of PCMDI/CMIP3 climate models in Pre-Industrial conditions.

| Models - PI scenario | $T_E^>$ $(K)$ | $T_E^<$ $(K)$ | $\eta_h$ | $\overline{F}/A$ $(Wm^{-2})$ | $\overline{W}_{min}/A$ $(Wm^{-2})$ | $Be_{max}$ | $\overline{\dot{S}_{mat}^{hor}}/A$ $(WK^{-1}m^{-2})$ | $\overline{\dot{S}_{mat}^{vert}}/A$ $(WK^{-1}m^{-2})$ | $\overline{\dot{S}_{mat}}/A$ $(WK^{-1}m^{-2})$ |
|---|---|---|---|---|---|---|---|---|---|
| 1 - BCCR BCM2.0 | 241.9 | 257.6 | 0.061 | 20.6 | 1.25 | 10.0 | $5.2\times10^{-3}$ | $47.2\times10^{-3}$ | $52.4\times10^{-3}$ |
| 4 - CNRM CM3 | 241.5 | 256.2 | 0.057 | 19.2 | 1.10 [3.1]♦ | 11.7 [4.1]* | $4.6\times10^{-3}$ | $49.2\times10^{-3}$ | $53.8\times10^{-3}$ |
| 6 - CSIRO Mk3.5 | 243.8 | 254.5 | 0.042 | 21.7 | 0.92 | 14.5 | $3.8\times10^{-3}$ | $51.4\times10^{-3}$ | $55.2\times10^{-3}$ |
| 7 – FGOALS | 240.1 | 256.3 | 0.063 | 21.0 | 1.33 | 10.2 | $5.6\times10^{-3}$ | $52.0\times10^{-3}$ | $57.6\times10^{-3}$ |
| 8 - GFDL CM2.0 | 242.9 | 257.0 | 0.055 | 21.6 | 1.19 | 11.3 | $4.9\times10^{-3}$ | $50.5\times10^{-3}$ | $55.4\times10^{-3}$ |
| 9 - GFDL CM2.1 | 243.6 | 258.0 | 0.056 | 21.7 | 1.21 | 11.2 | $5.0\times10^{-3}$ | $51.4\times10^{-3}$ | $56.4\times10^{-3}$ |
| 10 - GISS AOM | 243.1 | 257.2 | 0.055 | 24.0 | 1.31 | 12.1 | $5.4\times10^{-3}$ | $60.4\times10^{-3}$ | $65.8\times10^{-3}$ |
| 13 - HAD CM3 | 243.5 | 259.0 | 0.060 | 21.6 | 1.29 [3.1]♣ | 10.0 [4.2]♠♦ [4.1]* | $5.3\times10^{-3}$ | $48.7\times10^{-3}$ | $54.0\times10^{-3}$ |
| 14 - HAD GEM | 243.4 | 259.8 | 0.063 | 22.6 | 1.43 | 9.4 | $5.9\times10^{-3}$ | $50.5\times10^{-3}$ | $56.4\times10^{-3}$ |
| 15 - INM CM3.0 | 243.5 | 255.5 | 0.047 | 23.9 | 1.13 | 12.7 | $4.6\times10^{-3}$ | $53.9\times10^{-3}$ | $58.5\times10^{-3}$ |
| 16 - IPSL CM4 | 243.5 | 258.1 | 0.057 | 24.9 | 1.41 | 9.5 | $5.8\times10^{-3}$ | $49.5\times10^{-3}$ | $55.3\times10^{-3}$ |
| 17 - MIROC3.2 hires | 243.4 | 257.3 | 0.054 | 20.5 | 1.11 | 12.1 | $4.5\times10^{-3}$ | $50.2\times10^{-3}$ | $54.7\times10^{-3}$ |
| 18 - MIROC3.2 medres | 242.3 | 255.9 | 0.053 | 21.4 | 1.14 [2.7]♦ | 11.2 [4.7]♠♦ | $4.7\times10^{-3}$ | $48.4\times10^{-3}$ | $53.1\times10^{-3}$ |
| 20 - ECHAM5/MPI OM | 242.8 | 256.8 | 0.055 | 24.6 | 1.34 [3.3]♦ [2.6]♥ | 10.4 [4.2]♠♦ [5.3]♠♥ | $5.6\times10^{-3}$ | $53.1\times10^{-3}$ | $58.7\times10^{-3}$ |

---

♦ Actual value of the intensity of the Lorenz Energy cycle in XX century simulation runs (Marquez et al. 2010).

* True Bejan number obtained from the entropy production diagnostics reported for the PI simulation run (Pascale et al. 2009).

♣ Actual value of the intensity of the Lorenz Energy cycle in the PI simulation run (Pascale et al. 2009).

♠ True Bejan number computed using Eq. (22) in the text using data of $\overline{W}_{min}$, $\overline{W}$, and $Be_{max}$.

* True Bejan number obtained from the entropy production diagnostics reported for the PI simulation run (Pascale et al. 2009).

♥ Actual value of the intensity of the Lorenz Energy cycle in the PI simulation run, but using lower resolution (Hernandez-Deckers and Von Storch 2010)

**Table 2:** Correlations between thermodynamic parameters of PCMDI/CMIP3 climate models in Pre-Industrial conditions. Statistically significant values are depicted in bold, where the half-width of the 95% confidence interval is 0.2.

| Models - PI scenario | $C(\dot{\overline{S}}_{mat}^{hor}, \eta_h)$ | $C(\dot{\overline{S}}_{mat}^{hor}, \overline{F})$ | $C(\overline{F}, \eta_h)$ | $C(\dot{\overline{S}}_{mat}^{hor}, \dot{\overline{S}}_{mat}^{vert})$ |
|---|---|---|---|---|
| 1 - BCCR BCM2.0 | 0.81 | **0.60** | 0.02 | **0.51** |
| 4 - CNRM CM3 | 0.86 | 0.19 | **-0.34** | **0.74** |
| 6 - CSIRO Mk3.5 | 0.99 | **-0.46** | **-0.59** | **0.59** |
| 7 - FGOALS | 0.82 | **0.38** | **-0.22** | **0.52** |
| 8 - GFDL CM2.0 | 0.90 | **-0.27** | **-0.65** | 0.06 |
| 9 - GFDL CM2.1 | 0.70 | 0.18 | **-0.58** | **0.24** |
| 10 - GISS AOM | 0.84 | 0.04 | **-0.51** | **-0.20** |
| 13 - HAD CM3 | 0.78 | **0.53** | -0.13 | **0.40** |
| 14 - HAD GEM | 0.93 | 0.04 | **-0.34** | 0.08 |
| 15 - INM CM3.0 | 0.96 | **0.36** | 0.07 | **0.55** |
| 16 - IPSL CM4 | 0.87 | 0.18 | **-0.33** | **0.61** |
| 17 - MIROC3.2 hires | 0.90 | **0.23** | **-0.23** | **0.24** |
| 18 - MIROC3.2 medres | 0.82 | **0.52** | -0.06 | **0.56** |
| 20 - ECHAM5/MPI OM | 0.88 | **0.29** | -0.19 | **0.69** |

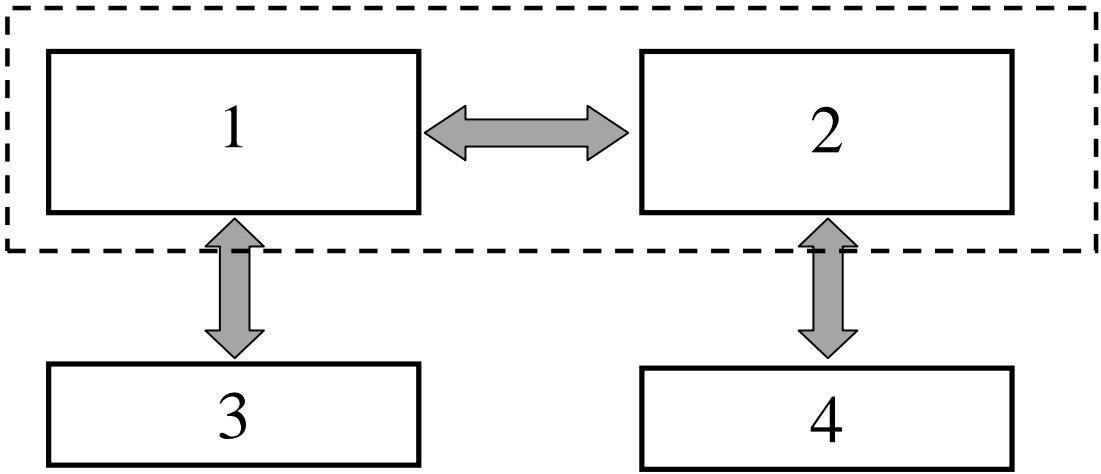

**Figure 1:** Minimal conceptual for the material entropy production of a planetary systems as deduced from Eq. (11). Boxes 1 and 2 represent warm (low latitudes) and cold (high latitudes) fluid domains, coupled by enthalpy transport. Boxes 3 and 4 represent warm (low latitudes) and cold (high latitudes) surface domains, coupled vertically to the corresponding fluid boxes, but not to each other. The dashed rectangle encloses the reduced two-box model usually considered in previous literature.

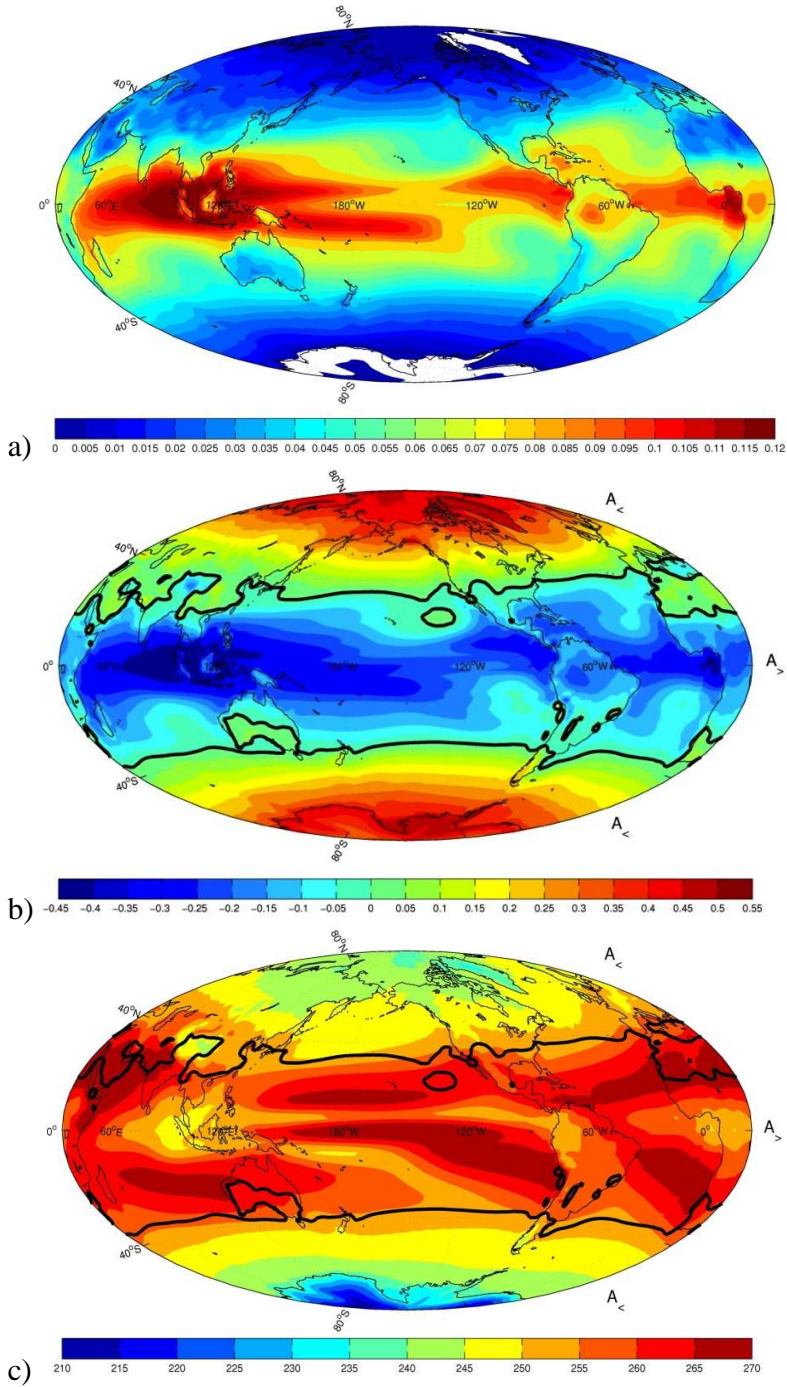

**Figure 2:** CM 13's time-averaged spatial fields of a) material entropy production via vertical processes $\overline{\dot{S}_{mat}^{vert}}$ (in $WK^{-1}m^{-2}$); b) material entropy production via horizontal processes $\overline{\dot{S}_{mat}^{hor}}$ ( in $WK^{-1}m^{-2}$); c) emission temperature (values in $K$). In b) and c) the solid black line separates the area $A_>$ with positive net energy budget at TOA from the area $A_<$ where the net budget is negative.

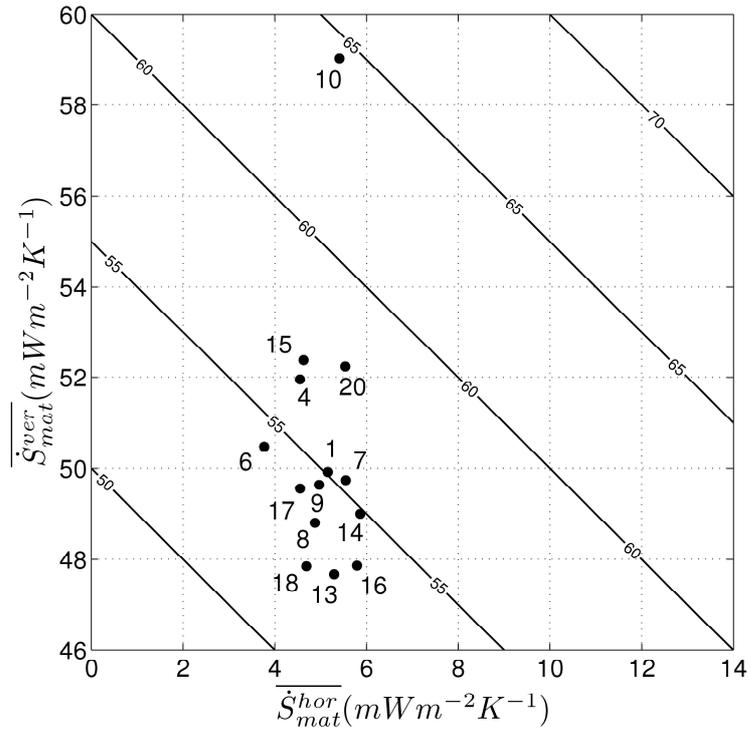

**Figure 3:** Material Entropy Production in PCMDI/CMIP3 climate models in the pre-industrial scenario runs. The contributions due to horizontal ($\overline{\dot{S}_{mat}^{hor}}$) and vertical ($\overline{\dot{S}_{mat}^{vert}}$) processes are depicted. Isolines of the total value of the material entropy production are shown as solid lines.

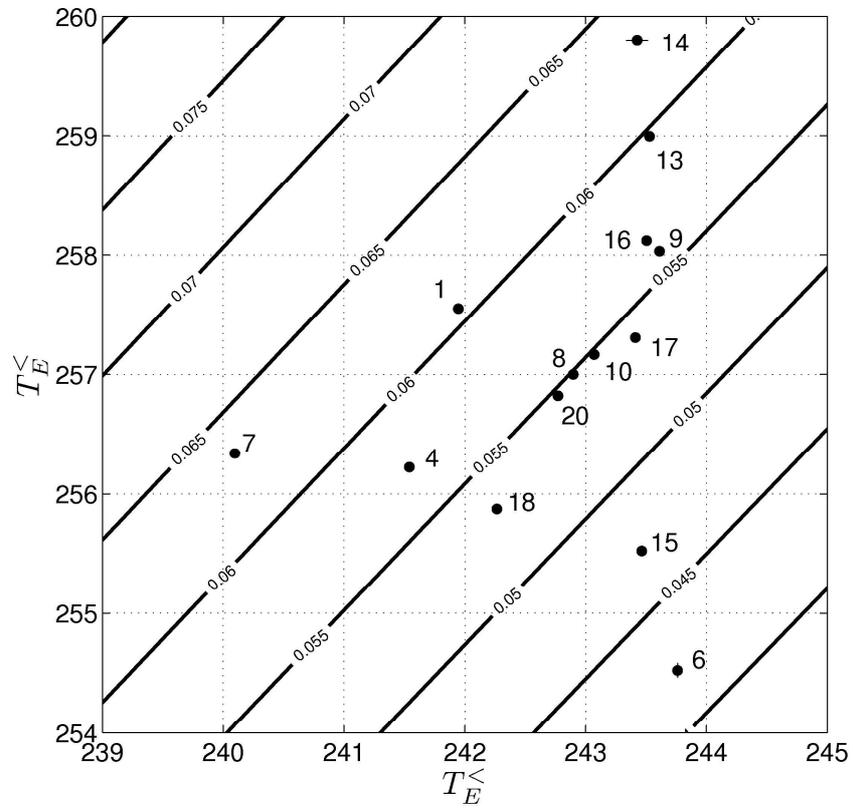

**Figure 4:** Equivalent temperatures of the warm ($T_E^>$) and cold ($T_E^<$) boxes of PCMDI/CMIP3 climate models in Pre-Industrial conditions. Isolines of efficiency $\eta_h$ are indicated with solid lines.

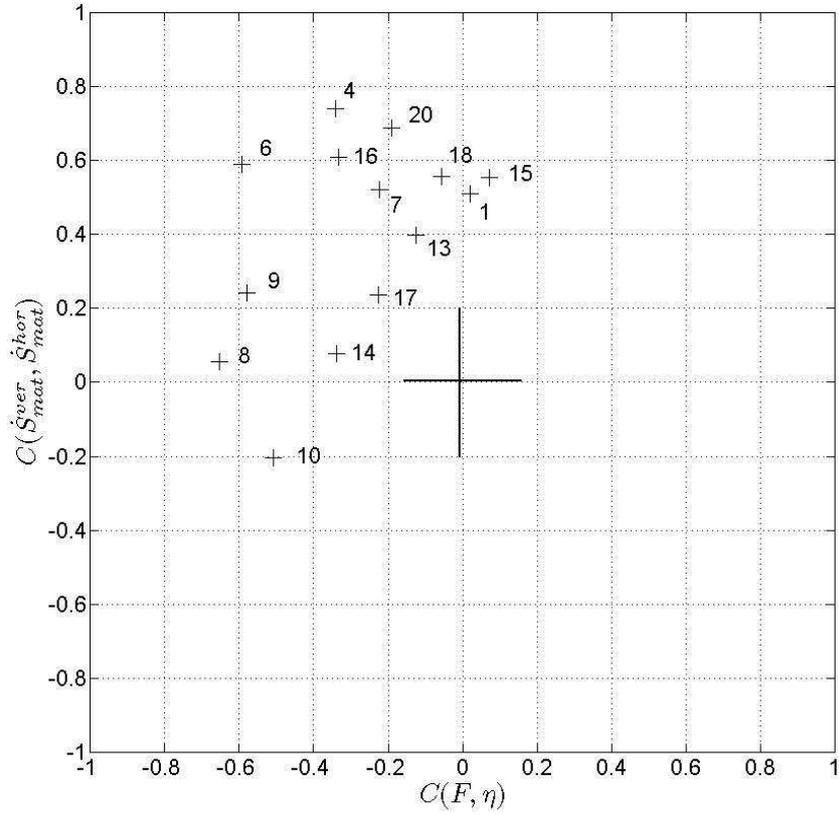

**Figure 5:** Strength of the large scale thermodynamic feedbacks for PCMDI/CMIP3 climate models in Pre-Industrial conditions. Correlation between total large scale heat transport $F$ and the baroclinic efficiency $\eta_h$ (x-axis) vs correlation between horizontal ($\overline{\dot{S}_{mat}^{hor}}$) and vertical $\overline{\dot{S}_{mat}^{ver}}$ components of the material entropy production (y-axis).

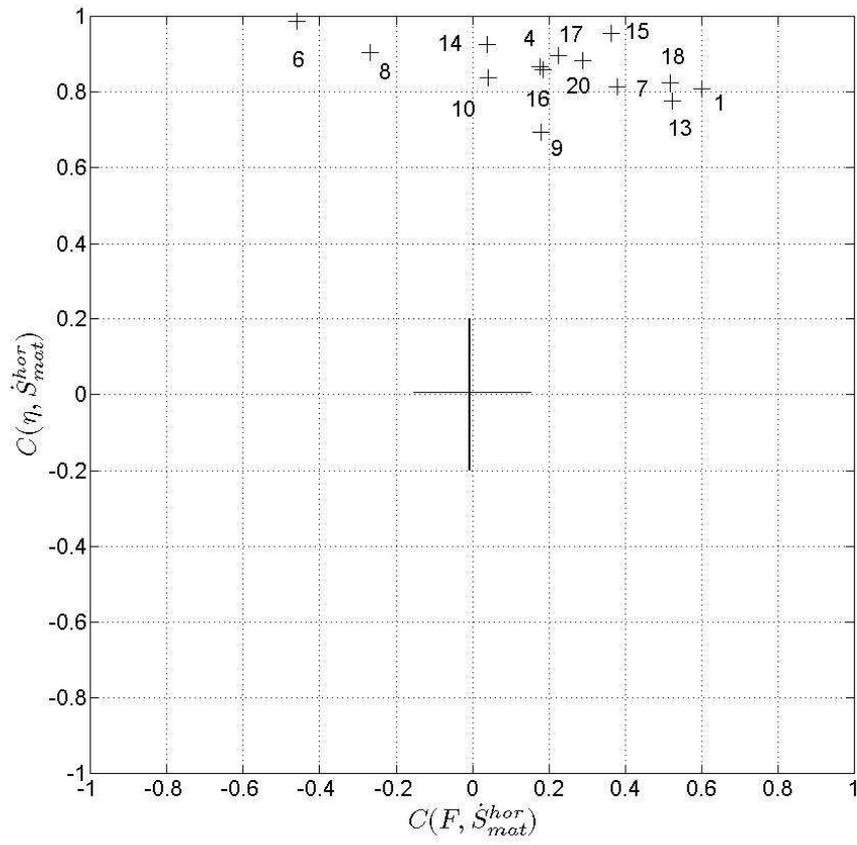

**Figure 6:** Correlations between horizontal component of material entropy production ($\overline{\dot{S}_{mat}^{hor}}$) and total meridional transport $\overline{F}$ (x-axis) and efficiency $\eta_h$ (y-axis).